%% file: R0_1.tex
\documentclass[journal,singleside,final]{IEEEtran}
\usepackage[nocompress]{cite}
\usepackage{amsmath}
\usepackage{amsthm}
\usepackage{amsfonts}
\usepackage{amssymb}
\usepackage[linesnumbered,ruled,vlined]{algorithm2e}
\usepackage{graphicx}
\usepackage{color}
\usepackage{xcolor}
\usepackage[hidelinks]{hyperref}
\usepackage[tight]{subfigure}
\usepackage{enumerate}
\usepackage{multirow}
\definecolor{orange}{rgb}{1,0.5,0}
\makeatletter
\newcommand{\removelatexerror}{\let\@latex@error\@gobble}
\makeatother

\SetKwInput{KwInput}{Input}                
\SetKwInput{KwOutput}{Output}              
\SetKw{Continue}{continue}
\SetKw{Break}{break}

\newcommand{\MATLAB}{\textsc{Matlab}\xspace}

\hypersetup{
    colorlinks=true,
    citecolor=blue,
    linkcolor=black,
    filecolor=magenta,      
    urlcolor=black,
    pdftitle={Overleaf Example},
    pdfpagemode=FullScreen,
    }

\input{abbrev.tex}

\renewenvironment{thebibliography}[1]{
  \begin{oldthebibliography}{#1}
    \setlength{\itemsep}{0.01em}
    \setlength{\parskip}{-0.12em}
}
{
  \end{oldthebibliography}
}

\begin{document}
\title{Beyond Diagonal RIS: Passive Maximum Ratio Transmission and Interference Nulling Enabler}
\author{Hamad~Yahya,~
\IEEEmembership{Member,~IEEE,} Hongyu~Li,~%
\IEEEmembership{Graduate~Student~Member,~IEEE,} Matteo~Nerini,~%
\IEEEmembership{Graduate~Student~Member,~IEEE,} Bruno~Clerckx,~%
\IEEEmembership{Fellow,~IEEE,} and~Merouane~Debbah,~%
\IEEEmembership{Fellow,~IEEE}
\thanks{Hamad Yahya is with the Department of Electrical Engineering, Khalifa University of Science and Technology, Abu Dhabi 127788, UAE (email: \href{mailto:hamad.myahya@ku.ac.ae}{hamad.myahya@ku.ac.ae}), he is also with the  Department of Electrical and Electronic Engineering, Imperial College London, London SW7 2AZ, U.K.

Hongyu Li, Matteo Nerini and Bruno Clerckx are with the  Department of Electrical and Electronic Engineering, Imperial College London, London SW7 2AZ, U.K., (e-mail: \{c.li21, m.nerini20, b.clerckx\}@imperial.ac.uk).

Merouane Debbah is with the 6G Research Center, Khalifa University of Science and Technology, Abu Dhabi 127788, UAE (email: \href{mailto:merouane.debbah@ku.ac.ae}{merouane.debbah@ku.ac.ae})
}}
\maketitle

\begin{abstract}
\Gls{bdris} generalizes and goes beyond conventional \gls{dris} by interconnecting elements to generate beyond diagonal scattering matrices, which significantly strengthen the wireless channels. In this work, we use \gls{bdris} for passive multiuser beamforming in \gls{mumiso} systems. Specifically, we design the scattering matrix of \gls{bdris} to either maximize the sum received signal power at the users following \gls{mrt}, or to nullify the interference at the users following \gls{zf}. To control the \gls{bdris} circuit topology complexity, we present the scattering matrix designs for the single/group/fully-connected \gls{bdris} architectures. Furthermore, we investigate uniform/optimized power allocation and \gls{zf} precoding at the \gls{bs}. Numerical results show that \gls{bdris} improves the interference nulling capability and sum rate with fewer \glspl{re} compared to \gls{dris}. In addition, at moderate to high \glspl{snr}, passive interference nulling reduces the complexity at the \gls{bs} by relaxing the need for precoding or water-filling power allocation design. Furthermore, the passive \gls{mrt} with \gls{zf} precoding achieves a tight sum rate performance to the joint design considering \gls{mumiso} scenarios with many \glspl{re} while maintaining low computational complexity and simplifying the channel estimation.
\end{abstract}
\glsresetall

\markboth{Draft,~Vol.~xx, No.~xx,
October~2024}{YAHYA \MakeLowercase{\textit{et al.}}: Beyond Diagonal RIS: Passive Maximum Ratio Transmission and Interference Nulling Enabler} 

\IEEEpeerreviewmaketitle
\glsresetall

\section{Introduction}\label{sec1}
\IEEEPARstart{I}{ntelligent} surfaces such as \gls{ris}, also known as \gls{irs}, are emerging technologies for the \gls{6g} mobile networks \cite{Marco2020-JSAC}, and have recently been introduced to complement active relays \cite{Marco2020-OJCOM,Jiancheng2022-TCOM}. Unlike active relays, \gls{ris} is nearly-passive and noiseless, only powering the controller which alters the electrical properties of the \glspl{re} by controlling the impedances attached to it \cite{Yuanwei2021-COMST}. Benefiting from its passiveness and reconfigurability, \gls{ris} has been recognized as a power-efficient enabler to engineer the wireless channel and enhance the communication performance by manipulating the phase and amplitude of the impinging signals without active circuitry and \gls{rf} chains \cite{Emil2022-MSP}. 
\subsection{Related Work}
\Gls{ris} technologies have gained significant attention recently due to their potential to enhance the communication performance of wireless systems. Among these technologies, two prominent categories have been intensively investigated in the literature, which are \gls{dris} \cite{Wu2019-TWC,Yang2021-TCOM,Nadeem2020-OJCOM,Liu2022-TWC,Jiang2022-JSAC,Zhu2024robust-Archive,Jangsher2023-TAES} and \gls{bdris} \cite{Shanpu2022-TWC,Hongyu2024-ComMag,MatteoGraph2024-TWC,Hongyu2022-TWC,Hongyu2023-JSAC,Rev3_1,Matteo2024-TWC,Rev3_2,Fang2024-CommLett,Rev2_1,Liu2024-LWC,chen2024transmitter,Li2022-TVT,Rev1_2,Rev1_1}. While these two categories are often discussed separately, it is important to note that \gls{bdris} can be viewed as a generalization of \gls{dris}. In other words, \gls{bdris} encompasses the characteristics of \gls{dris} as a special case, offering a broader and more flexible framework for reconfigurable intelligent surfaces. The key limitation in \gls{dris} is that it manipulates only the phase of the incident signal with $N$ reconfigurable impedances attached to $N$ \glspl{re} making single-connected circuit topology. In contrast, \gls{bdris} manipulates both the phase and amplitude of the incident signal with a network of reconfigurable impedances that connect the $N$ \glspl{re} together \cite{Shanpu2022-TWC,Hongyu2024-ComMag}. By interconnecting all the \glspl{re} to each other through $\frac{N(N+1)}{2}$ reconfigurable impedances, the fully-connected \gls{bdris} has been proposed, having the highest flexibility yet the highest circuit topology complexity. To balance the performance of the fully-connected \gls{bdris} with its circuit topology complexity, group-connected \gls{bdris} is introduced by Shen \textit{et al.} \cite{Shanpu2022-TWC}. In this architecture, the \glspl{re} are divided into $G$ groups, each with a group size of $N_G = \frac{N}{G}$ \glspl{re}. Within each group, the \glspl{re} are interconnected using $\frac{N_G(N_G+1)}{2}$ reconfigurable impedances, reducing the total number of reconfigurable impedances to $\frac{N(N_G+1)}{2}$. Additionally, graph theory is utilized to design optimal architectures with lower circuit topology complexity such as forest-connected and tree-connected \cite{MatteoGraph2024-TWC}. 

Focusing on various \gls{ris} architectures, in existing literature, beamforming design for both \gls{dris} \cite{Wu2019-TWC,Yang2021-TCOM,Nadeem2020-OJCOM,Liu2022-TWC,Jiang2022-JSAC,Zhu2024robust-Archive,Jangsher2023-TAES} and \gls{bdris} \cite{Hongyu2022-TWC,Hongyu2023-JSAC,Rev3_1,Matteo2024-TWC,Rev3_2,Fang2024-CommLett,Rev2_1,Liu2024-LWC,chen2024transmitter,Li2022-TVT,Rev1_2,Rev1_1} aided \gls{mumiso} systems have been studied and explained in detail below. 

\textbf{D-RIS}: The authors of \cite{Wu2019-TWC} minimize the transmit power at the \gls{bs} while maintaining a certain \gls{sinr} level at the users by jointly optimizing the scattering matrix of \gls{dris} and the active precoder at the \gls{bs} using the \gls{ao} algorithm. Similarly, Yang \textit{et al.} \cite{Yang2021-TCOM} achieve the same goal using passive \gls{mrt}/\gls{zf} at the \gls{dris} while iteratively optimizing the power allocation at the \gls{bs}. Considering the same setup, the authors of \cite{Nadeem2020-OJCOM} use \gls{ao} framework to maximize the minimum rate for a better user fairness. The authors of \cite{Liu2022-TWC} derive a closed-form steering vectors-based multiuser linear precoder to null the known interference and minimize the unknown interference. Jiang \textit{et al.} \cite{Jiang2022-JSAC} use the \gls{ao} framework to design the scattering matrix for interference nulling, achieving maximum degrees-of-freedom with high probability when the number of \glspl{re} is $N>2K(K-1)$, with $K$ the number of users. The authors of \cite{Zhu2024robust-Archive} present a robust gradient-based meta-learning approach without pre-training to design the scattering matrix of the \gls{dris} and the precoder  at the \gls{bs}. Jangsher \textit{et al.} \cite{Jangsher2023-TAES} maximize the sum rate of a wireless \gls{uav} with \gls{ris} considering energy efficiency constraints and practical limitations such as phase compensation and imperfect \gls{csi}.

\textbf{BD-RIS}: A joint design for the \gls{bdris} scattering matrix and \gls{bs} precoder is considered in \cite{Hongyu2022-TWC,Hongyu2023-JSAC,Rev3_1,Matteo2024-TWC,Rev3_2}, unlike \cite{Fang2024-CommLett} which considers a two-stage design. The authors of \cite{Hongyu2022-TWC,Hongyu2023-JSAC} maximize the sum rate of a multi-sector \gls{bdris} aided system by transforming the original problem into a multi-block optimization using auxiliary variables and solving it iteratively until convergence. In \cite{Rev3_1}, a unified approach is proposed for various objectives such as energy efficiency maximization, energy consumption minimization and sum rate maximization, while satisfying all constraints on the \gls{bdris} scattering matrix with additional quality-of-service constraints. In \cite{Matteo2024-TWC}, Nerini \textit{et al.} derive a closed-form \gls{bdris} scattering matrix that maximizes the channel gain of the single-user \gls{siso} scenario and the solution is extended to the \gls{mumiso} scenario. 
Sun \textit{et al.} \cite{Rev3_2} utilize the Takagi factorization-based symmetric unitary projection and \gls{ao} to jointly design the \gls{bs} precoder and the fully-connected \gls{bdris} scattering matrix. The \gls{bs} precoder aims to minimize the transmitted power and satisfy certain \gls{sinr} thresholds at the users, while the \gls{bdris} scattering matrix aims to maximize the equivalent channel's squared trace magnitude. Fang and Mao 
\cite{Fang2024-CommLett} consider \gls{bdris} with group/fully-connected architectures and derive a closed-form \gls{bdris} scattering matrix to maximize the sum of the users' equivalent channel gains following the gradient decent approach and the \gls{svd}-based symmetric unitary projection, while regularized \gls{zf} precoding is considered at the \gls{bs}.

Furthermore, \gls{bdris} has been investigated in cell-free massive \gls{mimo} \cite{Rev2_1}, and \gls{isac} systems \cite{Liu2024-LWC,chen2024transmitter}. For instance, the authors of \cite{Rev2_1} employed \gls{bdris} in \gls{swipt} systems to improve the wireless power transfer. The authors of \cite{Liu2024-LWC} jointly optimize the \gls{bs} linear filter, and precoder, as well as the \gls{bdris} scattering matrix to maximize the sum rate while satisfying the sensing requirements. Chen and Mao \cite{chen2024transmitter} propose a two-stage design to jointly maximize the sum rate and minimize the largest eigenvalue of the Cramér-Rao bound matrix for multiple sensing targets. While the aforementioned works consider reciprocal impedance networks, non-reciprocal single/group/multi-sector-connected \gls{bdris} architectures are proposed in \cite{Li2022-TVT,Rev1_2,Rev1_1} by relaxing the symmetry constraint.

\subsection{Motivations and Contributions}\label{sec1-motiv}
Although a local optimal joint design was proposed in \cite{Hongyu2022-TWC,Hongyu2023-JSAC}, the associated computational complexity of the proposed solution is significant. Therefore, a lower complexity two-stage design is proposed for multiuser systems in \cite{Fang2024-CommLett}. Nonetheless, the focus in \cite{Fang2024-CommLett} is primarily on designing the \gls{bdris} scattering matrix to maximize the norm of the equivalent channel gains while the inter-user interference is neglected, and thus results in performance loss. Hence, it is more suitable for the point-to-point systems as shown in \cite{Santamaria2023-SPL}. Therefore, this article focuses on efficiently and effectively addressing the interference management issue in multiuser systems by exploiting the interference management capability at the \gls{bdris}. The contributions of this work can be summarized as follows: 
\begin{enumerate}
    \item An efficient and low-complexity two-stage design is proposed for the \gls{bdris} scattering matrix and the \gls{bs} precoding. Specifically, we analyze the received \gls{sinr} at stage 1. We then design the \gls{bdris} scattering matrix to either maximize the sum received signal powers at the users following \gls{mrt} structure, known as \textit{passive \gls{mrt}}, or to nullify the interference following \gls{zf} structure, known as \textit{passive interference nulling}. The proposed \gls{bdris} scattering matrix design is general for the single/group/fully-connected \gls{bdris} architectures.
    \item Closed-form \gls{bdris} scattering matrix expressions are presented for the passive \gls{mrt} following the \gls{svd}-based symmetric unitary projection \cite{Fang2024-CommLett}, while an efficient \gls{ao} algorithm is presented for the passive interference nulling. It is found that applying the \gls{svd}-based symmetric unitary projection improves the received \gls{sinr} at stage 1 for the passive \gls{mrt}.
    \item For stage 2, we consider uniform/optimized power allocation and \gls{zf} precoding at the \gls{bs}. Specifically, the optimized power allocation aims to maximize the sum rate while the \gls{bdris} scattering matrix is given. For instance, passive \gls{mrt} requires optimization to achieve sum rate maximization, meanwhile when the passive interference nulling is adopted at the \gls{bdris}, water-filling power allocation is applied at the \gls{bs}. On the other hand, \gls{zf} precoding at the \gls{bs} is proposed for the passive \gls{mrt}.
    \item The computational complexity for both stages is analyzed considering the two \gls{bdris} scattering matrix designs, the three \gls{bdris} architectures, and the \gls{bs} processes. In addition, the channel state information requirements for each design are specified. For instance, it is found that the passive \gls{mrt} eases the channel state information requirements which further simplifies the channel estimation.
    \item The numerical results show that the \gls{bdris} enhances the interference nulling capability compared to the \gls{dris}. For instance, \gls{bdris} can support more users simultaneously for the same number of \glspl{re}. Also, the algorithm for \gls{bdris} converges rapidly with a smaller number of iterations and have a more stable interference nulling regardless of the initialization value. 
\end{enumerate}

\subsection{Article Organization}
The rest of the article is organized as follows. Sec. \ref{sec-system-model} introduces the system and channel models of the \gls{bdris}-enabled \gls{mumiso} systems. Sec. \ref{sec-stage1} is the \gls{bdris} scattering matrix design for passive \gls{mrt} and interference nulling considering single/group/fully-connected \gls{bdris} architectures. Sec. \ref{sec-stage2} is the \gls{bs} power allocation/precoding design. Sec. \ref{sec-results} is the numerical results and discussions. Finally, Sec. \ref{sec-concl} concludes the article with the main remarks and future work. Interested readers are referred to the \MATLAB code for this article which is available on \href{https://github.com/hmjasmi/bdris-mrt-null}{https://github.com/hmjasmi/bdris-mrt-null}.  

\subsection{Notations}
The notations used throughout the paper are explained as follows. Boldface uppercase and lowercase symbols such as $\mathbf{X}$ and $\mathbf{x}$ will denote matrix and column vectors, respectively. The matrix Kronecker product symbol is $\otimes$, the matrix Khatri-Rao product is $\odot$. The transpose, Hermitian transpose, complex conjugate and Moore-pseudoinverse are denoted by $(\cdot)^T$, $(\cdot)^H$, $(\cdot)^\ast$ and $(\cdot)^{\dag}$. Also, $(x)^+=\max(x,0)$, $\lceil\cdot\rceil$ is the ceiling function, $\mathbb{E}[\cdot]$ is the statistical expectation, $\mathrm{Re}\left[\cdot\right]$ is the real value of a complex number, $\vert\cdot\vert$ is the absolute value, $\Vert\cdot\Vert_2$ is the $l^2$ norm, $\Vert\cdot\Vert_F$ is the Frobenius norm, $\mathrm{Tr}(\cdot)$ is the trace, $[\mathbf{X}]_{i,j}$ is $i$-th row and $j$-th column entry in $\mathbf{X}$, $[\mathbf{X}]_{i:\bar{i},j:\bar{j}}$ extracts the $i$-th to $\bar{i}$-th rows and $j$-th to $\bar{j}$-th columns of $\mathbf{X}$, $[\mathbf{x}]_{i}$ is $i$-th entry in $\mathbf{x}$. The identity matrix and zeros vectors are denoted as $\textbf{I}$ and $\textbf{0}$, respectively. In addition, $\mathbf{x}=\mathrm{vec}\left(\mathbf{X}\right)$ is the matrix vectorization process which concatenate the columns of the matrix in one column vector, $\mathbf{X}=\mathrm{unvec}\left(\mathbf{x},[a,b]\right)$ reshapes the vector to an $a\times b$ matrix, $\mathrm{diag}\left(\mathbf{x}\right)$ is a diagonal matrix with the main diagonal being $\mathbf{x}$, $\mathrm{blkdiag}\left(\mathbf{X}_1,\mathbf{X}_2,\ldots,\mathbf{X}_G\right)$ is a $G\times G$ block matrix with the off-diagonal blocks being $\mathbf{0}$ and the diagonal blocks being $\mathbf{X}_1,\mathbf{X}_2,\ldots,\mathbf{X}_G$. The complex normal random variable with a zero mean and $\sigma^2$ variance is denoted as  $\mathcal{CN}(0,\sigma^2)$.

\section{System and Channel Models}\label{sec-system-model}
\begin{figure}[t]
    \centering
    \includegraphics[width=3.4in]{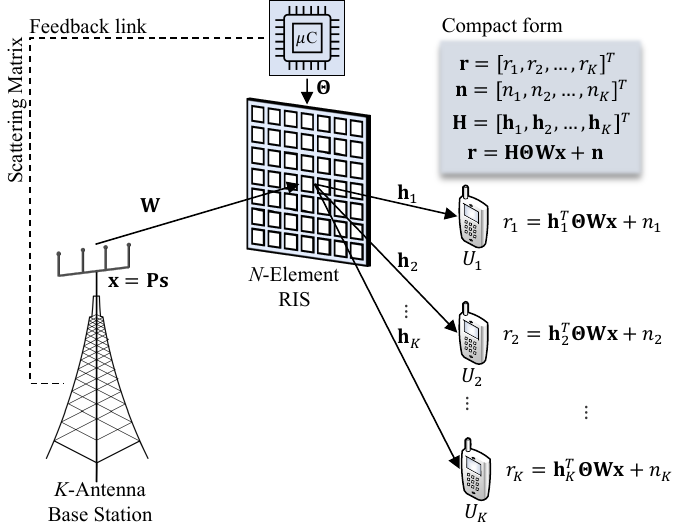}
    \caption{System block diagram.}
    \label{fig:BlockDiagram}
\end{figure}

In a downlink \gls{mumiso} system, the \gls{bs} is equipped with $K$ transmit antennas to serve $K$ single antenna users through a \gls{bdris} with $N$ \glspl{re}. The precoded signal vector to be transmitted at the antenna ports of the \gls{bs} can be expressed as $\mathbf{x}=\mathbf{Ps}$,
where $\mathbf{s}\in\mathbb{C}^{K \times 1}$ is the information symbols vector with $\mathbb{E}\left[\mathbf{ss}^H\right]=\mathbf{I}$, and the precoding matrix is denoted by $\mathbf{P}\in\mathbb{C}^{K\times K}$ such that $\left\Vert\mathbf{P}\right\Vert^2_F=P_{\max}$ with $P_{\max}$ the maximum transmission power at the \gls{bs}. 
Let the $k$th user be denoted as $U_k$, $k\in\{1,2,\ldots,K\}$, the received signal at $U_k$ is expressed as 
\begin{equation}
r_{k}=\mathbf{h}_{k}^{T}\boldsymbol{\Theta} \mathbf{Wx}+n_{k} \label{eq-rx}
\end{equation}
where $\mathbf{W}\in\mathbb{C}^{N\times K}$ is the \gls{bs}-\gls{bdris} channel matrix, $\boldsymbol{\Theta}\in\mathbb{C}^{N\times N}$ is the \gls{bdris} scattering matrix, $\mathbf{h}_k\in\mathbb{C}^{N\times 1}$ is \gls{bdris}-$U_k$ channel vector, and $n_{k}\sim\mathcal{CN}\left(0,N_0\right)$ is the \gls{awgn} with $N_0$ the power spectral density. It is worth noting that the \gls{bs}-$U_k$ channel is assumed to be blocked to highlight the performance gain of the \gls{bdris} \cite{Matteo2024-TWC}. To allow a systematic design of the scattering matrix, \eqref{eq-rx} can be written in a compact form such that the received signal vector is $\mathbf{r}=\mathbf{H}\boldsymbol{\Theta} \mathbf{Wx}+\mathbf{n}\in\mathbb{C}^{K\times 1}$, where  $\mathbf{H}=\left[\mathbf{h}_1,\ldots,\mathbf{h}_K\right]^T\in\mathbb{C}^{K \times N}$ is the concatenated \gls{bdris}-$U_k$, $\forall k$ channel matrix, and $\mathbf{n}$ is the \gls{awgn} vector with \gls{iid} entries. Fig. \ref{fig:BlockDiagram} depicts the system block diagram. 

Depending on the circuit topology for the \gls{bdris} reconfigurable impedances network, three architectures are considered \cite{Shanpu2022-TWC}:
\begin{enumerate}
    \item \underline{Single-connected \gls{bdris}:} The single-connected \gls{bdris} is equivalent to the conventional \gls{dris}. Each \gls{re} in this architecture is independently connected with a single grounded reconfigurable impedance without any interconnections among different \glspl{re}. Consequently, the scattering matrix of the single-connected \gls{bdris} is a diagonal matrix, which gives rise to the term \gls{dris} representing the independent operation of each \gls{re}. The constraint on the scattering matrix of the single-connected \gls{bdris} is given as
    \begin{equation}
        \mathcal{S}_{\mathrm{SC}_1}=\left\{\boldsymbol{\Theta}:\left[\boldsymbol{\Theta}\right]_{i,j}=0,\,\forall i\neq j\right\}
    \end{equation}
    where $i,j\in\{1,2,\ldots,N\}$. Considering purely reactive impedances, the scattering matrix of the lossless single-connected \gls{bdris} shall satisfy the unit-modulus constraint for all its elements, i.e., 
    \begin{equation}
        \mathcal{S}_{\mathrm{SC}_2}=\left\{\boldsymbol{\Theta}:\left\vert\left[\boldsymbol{\Theta}\right]_{i,j}\right\vert=1,\,\forall i= j\right\}\label{eq-unim-const}
    \end{equation}
    \item \underline{Fully-connected \gls{bdris}:} In the fully-connected \gls{bdris} architecture, each \gls{re} is interconnected with all other \glspl{re} through a network of reconfigurable impedances, allowing  waves impinging on one \gls{re} to be reflected by all other \glspl{re}. Consequently, the scattering matrix of the fully-connected \gls{bdris} is a full, reciprocal matrix due to the linearity of the reconfigurable impedances network. Therefore, the symmetry constraint on the scattering matrix of the fully-connected \gls{bdris} is given as
    \begin{equation}
    \mathcal{S}_{\mathrm{FC}_1}=\left\{\boldsymbol{\Theta}:\boldsymbol{\Theta}=\boldsymbol{\Theta}^T\right\}
    \end{equation}
     Considering purely reactive impedances, the scattering matrix of the lossless fully-connected \gls{bdris} shall satisfy the unitary constraint, i.e., 
    \begin{equation}
        \mathcal{S}_{\mathrm{FC}_2}=\left\{\boldsymbol{\Theta}:\boldsymbol{\Theta}\boldsymbol{\Theta}^H=\mathbf{I}\right\}
    \end{equation}

    \item \underline{Group-connected \gls{bdris}:}
    The group-connected \gls{bdris} architecture divides the \glspl{re} into $G$ groups, each with a group size of $N_G = \frac{N}{G}$ \glspl{re}. Within each group, the \glspl{re} are interconnected with all other intra-group \glspl{re} through a network of reconfigurable impedances, allowing  waves impinging on one \gls{re} to be reflected by all other intra-group \glspl{re} only because inter-group \glspl{re} are disconnected. Consequently, the scattering matrix of the group-connected \gls{bdris} is a block diagonal, with reciprocal block matrices due to the linearity of the reconfigurable impedances network in each group. Therefore, the symmetry constraint on each block of the scattering matrix of the group-connected \gls{bdris} is given as 
    \begin{multline}
        \mathcal{S}_{\mathrm{GC}_1}=\bigl\{\boldsymbol{\Theta}:\boldsymbol{\Theta}=\mathrm{blkdiag}\left(\boldsymbol{\Theta}_{1},\ldots,\boldsymbol{\Theta}_{G}\right),\, \\
        \boldsymbol{\Theta}_g=\boldsymbol{\Theta}_g^T,\,\forall g\bigr\}
    \end{multline}
where $\boldsymbol{\Theta}_g\in\mathbb{C}^{N_G\times N_G}$, $g\in\{1,2,\ldots,G\}$. Considering purely reactive impedances, the scattering matrix of the lossless group-connected \gls{bdris} shall satisfy the unitary constraint on each block $\boldsymbol{\Theta}_g$, i.e., 

    \begin{equation}
        \mathcal{S}_{\mathrm{GC}_2}=\bigl\{\boldsymbol{\Theta}:\boldsymbol{\Theta}_g\boldsymbol{\Theta}_g^H=\mathbf{I},\,\forall g\bigr\}
    \end{equation}
It is worth noting that when $G=1$ and $G=N$, the group-connected boils down to the fully-connected and single-connected architectures, respectively.

\end{enumerate}

While $\boldsymbol{\Theta}$ and $\mathbf{P}$ are given, the received \gls{sinr} at $U_k$ is
\begin{equation}
\gamma_k=\frac{\left\vert \mathbf{h}_{k}^{T}\boldsymbol{\Theta}\mathbf{W}\mathbf{p}_k\right\vert ^{2}}{\sum_{i\neq k}\left\vert \mathbf{h}_{k}^{T}%
\boldsymbol{\Theta}\mathbf{W}\mathbf{p}_i\right\vert ^{2}+N_0}
\end{equation}
Consequently, sum rate maximization problem is formulated as
\begin{subequations}
\begin{equation}
(P1): \,\max_{\mathbf{P},\,\boldsymbol{\Theta}}\sum_k\log_2\left(1+\gamma_k\right)\label{eq-main-opt}
\end{equation}
\begin{align}
\text{Subject to, } &&
\left\Vert\mathbf{P}\right\Vert^2_F&= P_{\max}\label{eq-fro-const}\\
&&\boldsymbol{\Theta}&\in\mathcal{S}_{c_1} \label{eq-recip-const}\\
&&\boldsymbol{\Theta}&\in\mathcal{S}_{c_2} \label{eq-loss-const}
\end{align}
\end{subequations}
where $c\in\{\mathrm{SC},\,\mathrm{FC},\,\mathrm{GC}\}$. 
The objective function shown in \eqref{eq-main-opt} is non-convex and shows a strong coupling between the precoding matrix and the scattering matrix, which makes it difficult to jointly optimize the two matrices. The constraint shown in \eqref{eq-fro-const} is non-convex. The constraint shown in \eqref{eq-recip-const} stems from the \gls{bdris} circuit topology, which is non-convex in general. The constraint shown in \eqref{eq-loss-const} is non-convex \cite{Hongyu2023-JSAC}, and it stems from the lossless nature of the \gls{bdris}. It is widely known as the Stiefel manifold condition \cite{Rev3_1}. Therefore, it is suggested to have a two-stage design to decouple the two optimization variables. Unlike \cite{Fang2024-CommLett}, our proposed design directly enhances the \gls{sinr} by applying passive \gls{mrt}/interference nulling at the \gls{bdris}. Such approach is motivated by the fact that utilizing \gls{zf} solely at the \gls{bs} usually incurs power penalty because most of the power is wasted for channel inversion, resulting in having very small power gains \cite{Zhang2009-TSP}. Hence, due to the \gls{bdris} passive nature, enhancing the \gls{sinr} by the \gls{bdris} is more effective than relying solely on the \gls{bs} for precoding.

\section{Stage 1: BD-RIS Scattering Matrix Design}\label{sec-stage1}
To explore to what extent \gls{bdris} could help to enhance the sum rate performance in \eqref{eq-main-opt}, we fix the impact of the \gls{bs} precoding by setting $\mathbf{P}=\sqrt{p_k}\mathbf{I}$, where $p_k={\frac{P_{\max}}{K}},\forall k$. Hence, the received signal vector can be written as 
\begin{equation}
\mathbf{r}=\sqrt{p_k}\mathbf{H}\boldsymbol{\Theta}\mathbf{Ws} + \mathbf{n} \triangleq \sqrt{p_k}\mathbf{H}\boldsymbol{\Omega}\mathbf{s} + \mathbf{n}  \triangleq \sqrt{p_k}\mathbf{Es} + \mathbf{n}
\end{equation}
where  $\boldsymbol{\Omega}=\boldsymbol{\Theta}\mathbf{W}=\left[\boldsymbol{\omega}_1,\ldots,\boldsymbol{\omega}_K\right]\in\mathbb{C}^{N \times K}$ and the equivalent channel is $\mathbf{E}=\mathbf{H}\boldsymbol{\Omega}=\left[\mathbf{e}_1,\ldots,\mathbf{e}_K\right]^T\in\mathbb{C}^{K\times K}$. Consequently, the received \gls{sinr} at $U_k$ stage 1 is given as
\begin{equation}
\gamma_k\!=\!\frac{p_k\left\vert \mathbf{h}_{k}^{T}\boldsymbol{\omega}_k\right\vert ^{2}}{p_k\sum_{i\neq k}\left\vert \mathbf{h}_{k}^{T}%
\boldsymbol{\omega}_i\right\vert ^{2}+N_0} 
\!=\! \frac{p_k\left\vert \left[\mathbf{e}_k\right]_{k}\right\vert ^{2}}{p_k\sum_{i\neq k}\left\vert \left[\mathbf{e}_k\right]_{i}\right\vert ^{2}+N_0}\label{eq-stage1-SINR}
\end{equation}
Considering \eqref{eq-stage1-SINR}, the rate at $U_k$ is influenced by the signal power that is determined by the numerator, as well as the interference which is determined by the summation in the denominator. Therefore, two objectives for the \gls{bdris} are motivated: 
\begin{enumerate}
    \item To maximize the desired signal power.
    \item To nullify the interference.
\end{enumerate}
The \gls{bdris} scattering matrix design following the first objective is called \textit{passive \gls{mrt}}, while the design following the latter is called \textit{passive interference nulling}. It is worth noting that maximizing $\left\Vert\mathbf{E}\right\Vert^2_F$ as in \cite{Fang2024-CommLett} would maximize $\sum_{k}\sum_{i}\left\vert[\mathbf{e}_k]_i\right\vert^2$. Hence, the numerator and denominator of \eqref{eq-stage1-SINR} are maximized simultaneously, which does not necessarily lead to the maximization of the \gls{sinr} in a \gls{mumiso} setting.
\subsection{Passive MRT}\label{subsec-MRT}
To maximize the desired signal power at all users simultaneously, the diagonal elements of $\mathbf{E}$ need to be considered, i.e., $\left[\mathbf{e}_k\right]_{k}$. Following the approach in \cite{Emil2014-MSP}, it is noted that $\boldsymbol{\omega}_k$ and $\exp{\left(j\theta_k\right)}\boldsymbol{\omega}_k$ for any common phase rotation $\theta_k\in\mathbb{R}$ lead to the same values of $|[\mathbf{e}_k]_k|$. Hence, the phase ambiguity can be exploited to make the inner product $\mathbf{h}_k^T\boldsymbol{\omega}_k$ real and positive without loss of optimality. This motivates us to maximize the real part of $\mathbf{h}_k^T\omega_k$ instead of its absolute value. Therefore, to capture the desired signal at all users, the trace of the real-valued equivalent channel will be used while relaxing the challenging constraints in \eqref{eq-recip-const} and \eqref{eq-loss-const}. Consequently, we can find a relaxed $\boldsymbol{\Theta}$ by solving the following problem
\begin{subequations}
\begin{align}
    (P2a): & \,\max_{\boldsymbol{\Theta}}f\left (\boldsymbol{\Theta}\right)\!\triangleq\!\mathrm{Tr}\left(\mathrm{Re}\left[\mathbf{E}\right]\right)\!=\!\mathrm{Tr}\left(\mathrm{Re}\left[\mathbf{H}\boldsymbol{\Theta}\mathbf{W}\right]\right) \label{eq-opt-maxSNR}\\
    &\text{Subject to, } \left\Vert\boldsymbol{\Theta}\right\Vert^2_F=N\label{eq-opt-maxSNR-const}
\end{align}
\end{subequations}
The objective function can be written as $f\left (\boldsymbol{\Theta}\right)=\mathrm{Tr}\left(\mathrm{Re}\left[\mathbf{E}\right]\right)=\frac{1}{2}\left(\mathrm{Tr}\left(\mathbf{E}\right)+\mathrm{Tr}\left(\mathbf{E}^{\ast}\right)\right)=\frac{1}{2}\left(\mathrm{Tr}\left(\mathbf{E}\right)+\mathrm{Tr}\left(\mathbf{E}\right)^{\ast}\right)=\mathrm{Re}\left[\mathrm{Tr}\left(\mathbf{E}\right)\right]$. Hence, $f\left (\boldsymbol{\Theta}\right)=\mathrm{Tr}\left(\mathrm{Re}\left[\mathbf{H}\boldsymbol{\Theta}\mathbf{W}\right]\right)=\mathrm{Re}\left[\mathrm{Tr}\left(\mathbf{H}\boldsymbol{\Theta}\mathbf{W}\right)\right]$. Using the cyclic property of the trace, we can write the objective function in \eqref{eq-opt-maxSNR} as $f\left (\boldsymbol{\Theta}\right)=\mathrm{Re}\left[\mathrm{Tr}\left(\mathbf{G}\boldsymbol{\Theta}\right)\right]$, where $\mathbf{G}=\mathbf{WH}\in\mathbb{C}^{N\times N}$ is the cascaded \gls{bs}-\gls{bdris}-$U_k$ channel matrix. In the following, we outline the passive \gls{mrt} solution for the single-connected and group-connected \gls{bdris} architectures, where the latter captures the fully-connected architecture as well.
\subsubsection{Solution for Single-Connected \gls{bdris}}
While noting that the single-connected architecture has a diagonal $\boldsymbol{\Theta}$ with unit modulus elements, we can expand $(P2a)$ as
\begin{subequations}
\begin{align}
    (P2a): & \,\max_{\boldsymbol{\Theta }}f\left( \boldsymbol{\Theta }\right) = \mathrm{Re}\left[
\sum_{i=1}^{N}\left[ \mathbf{G}\right] _{i,i}\left[ 
\boldsymbol{\Theta }\right] _{i,i} \right]
 \label{eq-opt-maxSNR-SC}\\
    &\text{Subject to, } \sum_{i=1}^{N}\left\vert \left[ \boldsymbol{\Theta }%
\right] _{i,i}\right\vert ^{2}=N
\end{align}
\end{subequations}
Therefore, the optimal scattering matrix that belongs to the feasible set, i.e., $\mathcal{S}_{\mathrm{SC}_1}$ and $\mathcal{S}_{\mathrm{SC}_2}$, is given as 
\begin{equation}
    \left[\boldsymbol{\Theta}^\star\right]_{i,j} = \left\{\begin{array}{lr}
        \Pi_{\mathcal{S}_\mathrm{unim}}\left(\left[\mathbf{G}\right]_{i,j}^{\ast}\right) ,& i=j\\
        0 ,& i\neq j
        \end{array}
        \right.\label{eq-mrt-final-SC}
\end{equation}
where \begin{equation}
\Pi_{\mathcal{S}_\mathrm{unim}}\left(\left[\mathbf{G}\right]_{i,j}^{\ast}\right)=\frac{\left[\mathbf{G}\right]_{i,j}^{\ast}}{\left\vert\left[\mathbf{G}\right]_{i,j}^{\ast}\right\vert}\label{eq:proj-unim}
\end{equation}
is the projection to the unit-modulus constraint shown in \eqref{eq-unim-const} with the minimum Euclidean distance to $\left[\mathbf{G}\right]_{i,j}^{\ast}$ \cite{Jiang2022-JSAC}. 

\textit{Computational Complexity}: The computational complexity of \eqref{eq-mrt-final-SC} depends on the computational complexity of computing the diagonal elements of $\mathbf{G}^{\ast}$, $\mathcal{O}\left(KN\right)$, while the normalization has a computational complexity $\mathcal{O}\left(N\right)$. Hence, the overall complexity can be written as $\mathcal{O}\left(KN+N\right)=\mathcal{O}\left(KN\right)$. 

\subsubsection{Solution for Group-Connected \gls{bdris}}
While noting that the group-connected architecture has a block diagonal $\boldsymbol{\Theta}$ with unitary $\boldsymbol{\Theta}_g$, we define ${\mathbf{G}}_g\triangleq\left[\mathbf{G}\right]_{N_g(g-1)+1:gN_g,N_g(g-1)+1:gN_g}\in\mathbb{C}^{N_g\times N_g}$. Hence, we can expand $(P2a)$ as
\begin{subequations}
\begin{align}
    (P2a): & \,\max_{\boldsymbol{\Theta }}f\left( \boldsymbol{\Theta }\right) = \mathrm{Re}\left[
\sum_{g=1}^{G}\sum_{i=1}^{N_g}\sum_{j=1}^{N_g}[ {\mathbf{G}}_g] _{i,j}\left[ 
\boldsymbol{\Theta }_g\right] _{j,i} \right]
 \label{eq-opt-maxSNRa-GC}\\
    &\text{Subject to, } \sum_{i=1}^{N_g}\sum_{j=1}^{N_g}\left\vert \left[ \boldsymbol{\Theta }_g%
\right] _{j,i}\right\vert ^{2}=N_g, \, \forall g
\end{align}
\end{subequations}
Consequently, the objective function in \eqref{eq-opt-maxSNRa-GC} can be maximized by co-phasing the terms $[ {\mathbf{G}}_g] _{i,j}$ and $\left[ 
\boldsymbol{\Theta }_g\right] _{j,i}$ before summation. Such problem has a closed-form solution that follows the \gls{mrt} precoding \cite{Lo1999-TCom} and the pseudo match-and-forward precoding for non-regenerative \gls{mimo} relays \cite{Tang2007-TWC}. Consequently, the closed-form solution given as 
\begin{align}
\boldsymbol{\Theta}&=\mathrm{blkdiag}\left(\boldsymbol{\Theta}_{1},\ldots,\boldsymbol{\Theta}_{G}\right) \notag \\
\boldsymbol{\Theta}_g&= {\mathbf{G}}_g^H\sqrt{%
\frac{N_g}{\mathrm{Tr}\left( {\mathbf{G}}_g{\mathbf{G}}_g^{H}\right) }}\label{eq:mrt-simple-GC}
\end{align}
Since \eqref{eq:mrt-simple-GC} does not belong to the feasible set, i.e., $\mathcal{S}_{\mathrm{GC}_1}$ and $\mathcal{S}_{\mathrm{GC}_2}$, we project such low-complexity $\boldsymbol{\Theta}$ to the feasible set using the \gls{svd}-based symmetric unitary projection. In other words, we formulate an optimization problem to find $\boldsymbol{\Theta}$ which belongs to the feasible set by minimizing the squared Frobenius distance between the low-complexity $\boldsymbol{\Theta}_g$ and the desired $\boldsymbol{\Theta}_g$. Hence, 
\begin{subequations}
\begin{align}
    (P2b): & \,\boldsymbol{\Theta}_g^\star = \arg \min_{\boldsymbol{\Theta}_g}\left\Vert \boldsymbol{\Theta}_g - \mathbf{G}_g^H\sqrt{%
\frac{N_g}{\mathrm{Tr}\left( \mathbf{G}_g\mathbf{G}_g^{H}\right) }} \right\Vert^2_F \\
    &\text{Subject to, } \eqref{eq-recip-const},\eqref{eq-loss-const}
\end{align}
\end{subequations}
Such approach is known as a matrix approximation and is used due to its ease of computation. In addition, the squared Frobenius norm is used specifically because it is reported that it maximizes the entropy for quantized channels in \gls{mimo} systems \cite{Bjornson2009-TSP}. This problem can be solved efficiently following \cite[Eq. (10)-(12)]{Fang2024-CommLett}, which projects the low-complexity solution to the feasible set, i.e., \eqref{eq:mrt-simple-GC} needs symmetric unitary projection. A closed-form  symmetric projection of $\boldsymbol{\Theta}_g$ is given as \cite[Eq. (10)]{Fang2024-CommLett},
\begin{equation}
\Pi_{\mathcal{S}_\mathrm{sym}}\left(\boldsymbol{\Theta}_g\right) = 0.5\left(\boldsymbol{\Theta}_g+\boldsymbol{\Theta}_g^T\right)\label{eq:proj-resp}
\end{equation}
The unitary projection of $\boldsymbol{\Theta}_g$ is given as \cite[Eq. (11)]{Fang2024-CommLett}, 
\begin{equation}
\Pi_{\mathcal{S}_\mathrm{uni}}\left(\boldsymbol{\Theta}_g\right) = \mathbf{UV}^H \label{eq:proj-loss}
\end{equation}
where its \gls{svd} can be written as $\boldsymbol{\Theta}_g=\mathbf{U}\boldsymbol{\Sigma}\mathbf{V}^H$, 
$\mathbf{U}$ and $\mathbf{V}$ are the unitary left and right singular vectors and $\boldsymbol{\Sigma}$ contains the singular values ordered in a descending order. Finally, the \gls{svd}-based symmetric unitary projection of $\boldsymbol{\Theta}_g$ is given as \cite[Eq. (12)]{Fang2024-CommLett},
\begin{equation}
\Pi_{\mathcal{S}_\mathrm{symuni}}\left(\boldsymbol{\Theta}_g\right) = \widehat{\mathbf{U}}\mathbf{V}^H \label{eq:proj-symuni}
\end{equation}
where $\Pi_{\mathcal{S}_\mathrm{sym}}\left(\boldsymbol{\Theta}_g\right)$ is a rank deficient matrix with rank $R$ and its unitary left and right singular vectors are
$\mathbf{U}=\left[\mathbf{U}_R,\mathbf{U}_{N-R}\right]$ and $\mathbf{V}=\left[\mathbf{V}_R,\mathbf{V}_{N-R}\right]$. Also, $\widehat{\mathbf{U}}=\left[\mathbf{U}_R,\mathbf{V}^{\ast}_{N-R}\right]$. Hence, we can express the optimal scattering matrix in closed-form as 
\begin{align}
\boldsymbol{\Theta}^\star &= \mathrm{blkdiag}\left(\boldsymbol{\Theta}^\star_{1},\ldots,\boldsymbol{\Theta}^\star_{G}\right) \label{eq:mrt-final-GC1} \\
\boldsymbol{\Theta}_g^\star&=\Pi_{\mathcal{S}_\mathrm{symuni}}\left(\mathbf{G}_g^H\sqrt{%
\frac{N_g}{\mathrm{Tr}\left( {\mathbf{G}_g}\mathbf{G}_g^{H}\right) }}\right)\label{eq:mrt-final-GC}
\end{align}

It should be noted that the signal in general will be degraded due to the \gls{svd}-based symmetric unitary projection. It will be seen in the Numerical Results and Discussions Section that the \gls{svd}-based symmetric unitary projection not only will degrade the signal power but also will reduce the interference power in \eqref{eq-stage1-SINR}.

\textit{Computational Complexity}: The computational complexity of \eqref{eq:mrt-final-GC} depends on the complexity of \eqref{eq:mrt-simple-GC}, $\mathcal{O}\left(GKN_g^2+GN_g^2\right)$, and is dominated by the $G$ \gls{svd} computations for \eqref{eq:proj-symuni}, $\mathcal{O}\left(GN_g^3\right)$. Hence, the overall complexity can be written as $\mathcal{O}\left(GN_g^3+GKN_g^2+GN_g^2\right)=\mathcal{O}\left(GN_g^3\right)$.

\textit{Channel State Information Requirements}: Furthermore, the cascaded \gls{bs}-\gls{bdris}-$U_k$ channel matrix, $\mathbf{G}$, is needed to compute $\boldsymbol{\Theta}^\star$ for the three architectures. This can be found by selecting the inter-impedance elements of the \gls{bdris} to have an infinitely large reactance. Hence, $\mathbf{G}$ can be estimated by setting $\boldsymbol{\Theta}=\mathbf{I}$ \cite{Demir2022-TWC}. Therefore, the channel estimation is simplified without separately estimating $\mathbf{H}$ and $\mathbf{W}$.

\subsection{Passive Interference Nulling}\label{subsec-intNull}
To nullify the interference terms in \eqref{eq-stage1-SINR}, the equivalent channel should result in a diagonal matrix, i.e., $\mathbf{E}=\mathbf{H}\boldsymbol{\Theta}\mathbf{W}=\mathrm{diag}\left(\boldsymbol{\lambda}\right)$ and $\boldsymbol{\lambda}\in\mathbb{C}^{K\times 1}$ is the interference-free equivalent channels at the users. Considering the most flexible architecture which is the fully-connected, an intuitive approach to design a low-complexity $\boldsymbol{\Theta}$ that nulls the interference follows the \gls{zf} precoding. Hence,
\begin{equation}    
\boldsymbol{\Theta}=\mathbf{G}^{\dag}\sqrt{%
\frac{N}{\mathrm{Tr}\left( \mathbf{G}^{\dag H}\mathbf{G}^{\dag}\right) }}\label{eq:zf-simple}
\end{equation} 
where $\mathbf{G}^{\dag}=\mathbf{H}^H\left(\mathbf{H}\mathbf{H}^H\right)^{-1}\left(\mathbf{W}^H\mathbf{W}\right)^{-1}\mathbf{W}^H$ and the normalization factor is to satisfy the constraint in \eqref{eq-opt-maxSNR-const}. Nonetheless, \eqref{eq:zf-simple} does not belong to the feasible set. Hence, the optimal symmetric unitary scattering matrix that is the closest to \eqref{eq:zf-simple} in terms of Frobenius norm can be written in closed-form as $\boldsymbol{\Theta}^\star = \Pi_{\mathcal{S}_\mathrm{symuni}}\left(\mathbf{G}^{\dag}\sqrt{%
\frac{N}{\mathrm{Tr}\left( \mathbf{G}^{\dag H}\mathbf{G}^{\dag}\right) }}\right)$. 

It is worth noting that the approximation error has a determinantal effect on the equivalent channel as $\mathbf{E}$ will be no longer diagonal which is proved in the following. 
\begin{proof}
While dropping the normalization factor in \eqref{eq:zf-simple}, the equivalent channel can be written as 
\begin{equation}
\mathbf{E}=\mathbf{HH}^\dag\mathbf{W}^\dag\mathbf{W}=\mathbf{I}
\end{equation}
While we consider the symmetry condition alone to facilitate the proof, the equivalent channel becomes 
\begin{align}
\widehat{\mathbf{E}}&=0.5\mathbf{HH}^\dag\mathbf{W}^\dag\mathbf{W}+0.5\mathbf{HW}^{\dag T}\mathbf{H}^{\dag T}\mathbf{W}\notag\\
&=0.5\mathbf{I}+0.5\mathbf{HW}^{\dag T}\mathbf{H}^{\dag T}\mathbf{W}
\end{align}
which is in general not a diagonal matrix. 
\end{proof}

To effectively nullify the interference, we adopt an alternative approach as detailed below. The adopted approach is inspired by the interference nulling presented for \gls{dris} in \cite{Jiang2022-JSAC}. Hence, to null the interference in a \gls{bdris}-enabled \gls{mumiso} system, the equivalent channel, i.e., $\mathbf{E}$, needs to be classified into a desired/interference channels. Then, $\boldsymbol{\Theta}$ shall be tuned to null all interference channels. In the following, we illustrate interference nulling for the group-connected architecture which captures the fully-connected and single-connected \gls{bdris} architectures as a special case.

We can re-write the equivalent channel as 
\begin{equation}
\mathrm{vec}\left(\mathbf{E}\right)=\mathbf{A}_c\boldsymbol{\theta}_c
\end{equation} 
where $\boldsymbol{\theta}_c\in\mathbb{C}^{GN_g^2\times 1}$ and is given as
\begin{equation}
    \boldsymbol{\theta}_c= \left\{\begin{array}{lc}
        \left[[\boldsymbol{\Theta}]_{1,1},[\boldsymbol{\Theta}]_{2,2},\ldots,[\boldsymbol{\Theta}]_{N,N}\right]^T,& c=\mathrm{SC}\\
        \mathrm{vec}(\boldsymbol{\Theta}),& c=\mathrm{FC} \\
        \left[\mathrm{vec}(\boldsymbol{\Theta}_1)^T,\ldots,\mathrm{vec}(\boldsymbol{\Theta}_G)^T\right]^T,& c=\mathrm{GC}
        \end{array}
        \right.\label{eq-theta-c}
\end{equation}
$\boldsymbol{\theta}_{\mathrm{SC}}\in\mathbb{C}^{N\times 1}$, $\boldsymbol{\theta}_{\mathrm{FC}}\in\mathbb{C}^{N^2\times 1}$, $\boldsymbol{\theta}_{\mathrm{GC}}\in\mathbb{C}^{GN_g^2\times 1} $, $\mathbf{A}_c\in\mathbb{C}^{K^2\times GN_g^2}$ denotes the desired/interference channels and is given as
\begin{equation}
\mathbf{A}_c=\left\{\begin{array}{lc}
        \mathbf{W}^T\odot\mathbf{H},& c=\mathrm{SC}\\
        \mathbf{W}^T\otimes\mathbf{H},& c=\mathrm{FC} \\
        \left[\mathbf{W}_1^T\otimes\mathbf{H}_1,\ldots,\mathbf{W}_G^T\otimes\mathbf{H}_G\right],& c=\mathrm{GC}
        \end{array}
        \right.
\end{equation} ${\mathbf{W}}_g=\left[\mathbf{W}\right]_{N_g(g-1)+1:gN_g,1:K}\in\mathbb{C}^{N_g\times K}$, ${\mathbf{H}}_g=\left[\mathbf{H}\right]_{1:K,N_g(g-1)+1:gN_g}\in\mathbb{C}^{K\times N_g}$, $\mathbf{A}_{\mathrm{SC}}\in\mathbb{C}^{K^2\times N}$, $\mathbf{A}_{\mathrm{FC}}\in\mathbb{C}^{K^2\times N^2}$, $\mathbf{A}_{\mathrm{GC}}\in\mathbb{C}^{K^2\times GN_g^2}$. To define the desired/interference channels, we drop the subscript defining the \gls{bdris} architecture for notational simplicity. For instance, $\mathbf{a}^T_i=[\mathbf{A}_c]_{i,:}$ represents the desired/interference channels for a specific user. We define $\tilde{\mathbf{a}}_{\grave{k},k}=\mathbf{a}_i$, for $i=(k-1)\times K + \grave{k}$. If $\grave{k}=k$, then $\tilde{\mathbf{a}}_{\grave{k},k}$ is the desired channel for $U_k$. Otherwise, $\tilde{\mathbf{a}}_{\grave{k},k}$ is the interference channel for $U_k$ from antenna $\grave{k}$. Consequently, the interference channels matrix can be defined as
\begin{subequations}
\begin{align}
\bar{\mathbf{A}}_c&=\left[\bar{\mathbf{A}}_1,\bar{\mathbf{A}}_2,\ldots,\bar{\mathbf{A}}_K\right]\in\mathbb{C}^{GN_g^2\times K(K-1)}\label{eq-intCH}\\
\bar{\mathbf{A}}_1 &=\left[\tilde{\mathbf{a}}_{2,1},\tilde{\mathbf{a}}_{3,1},\ldots,\tilde{\mathbf{a}}_{K,1}\right] \\
\bar{\mathbf{A}}_k &= [\tilde{\mathbf{a}}_{1,k},\cdots,\tilde{\mathbf{a}}_{k-1,k},\tilde{\mathbf{a}}_{k+1,k},\ldots,\tilde{\mathbf{a}}_{K,k}],  \notag\\
& \,\,\,\,\,\,\,\,\,1<k\leq K-1\\
\bar{\mathbf{A}}_K&=\left[\tilde{\mathbf{a}}_{1,K},\tilde{\mathbf{a}}_{2,K},\ldots,\tilde{\mathbf{a}}_{K-1,K}\right]\label{eq-intCHd}
\end{align}
\end{subequations}
In addition, the desired channels matrix is defined as $\check{\mathbf{A}}_c=\left[\tilde{\mathbf{a}}_{1,1},\tilde{\mathbf{a}}_{2,2},\ldots,\tilde{\mathbf{a}}_{K,K}\right]\in\mathbb{C}^{GN_g^2\times K}$.


To achieve $\mathbf{E}=\mathrm{diag}(\boldsymbol{\lambda})\Leftrightarrow\mathbf{A}_c\boldsymbol{\theta}=\boldsymbol{\lambda}$, the $K(K-1)$ interference channels needs to be nulled by the \gls{bdris}. This can be achieved by satisfying the nulling condition, which is expressed as
\begin{equation}
\bar{\mathbf{A}}_c^T\boldsymbol{\theta}_c=\mathbf{0} \label{eq-null}
\end{equation} 
In general, \eqref{eq-null} can be seen as a system of linear equations with $K(K-1)$ complex equations and $GN_g^2$ complex variables. Nonetheless, the number of effective variables is less due to the constraints on $\boldsymbol{\Theta}$, i.e., \eqref{eq-recip-const}--\eqref{eq-loss-const}, which is $\frac{N(1+N_g)}{2}$ real variables. Therefore, given that the channel realizations are random, \eqref{eq-null} can be achieved with high probability if the number of real variables is greater or equal the number of real equations, i.e., 
\begin{equation}
    \frac{N(1+N_g)}{2}\geq 2K(K-1) \label{eq-NoREs-null}
\end{equation}
which can be solved for $N$ considering $K>1$ as follows
\begin{equation}
    \left\{\begin{array}{lc}
        N\geq2K(K-1),& c=\mathrm{SC}\\
        N\geq2K-1,& c=\mathrm{FC} \\
        N\geq\left\lceil\frac{4K(K-1)}{1+N_g}\right\rceil,& c=\mathrm{GC}
        \end{array}
        \right.\label{eq-N-bound-null}
\end{equation}
The expression for the fully-connected architecture can be proved by solving the quadratic inequality and noting that $K$ and $N$ are integers, and $N_g=N$ showing that $\frac{-1+\sqrt{16K^2-16K+1}}{2}$ lies strictly between $2K-2$ and $2K-1$. Furthermore, \eqref{eq-NoREs-null} can be solved for $K$ as follows
\begin{equation}\renewcommand{\arraystretch}{1.5}
    \left\{{\begin{array}{lc}
        K\leq\left\lfloor\frac{1+\sqrt{1+2N}}{2}\right\rfloor,& c=\mathrm{SC}\\
        K\leq\left\lfloor\frac{N+1}{2}\right\rfloor,& c=\mathrm{FC} \\
        K\leq\left\lfloor\frac{1+\sqrt{1+N(1+N_g)}}{2}\right\rfloor,& c=\mathrm{GC}
        \end{array}}
        \right.
\end{equation}\renewcommand{\arraystretch}{1}

Consequently, the objective is to find $\boldsymbol{\Theta}$ that is subject to the interference nulling and \gls{bdris} constraints. In other words, we have the following feasibility-check problem
\begin{subequations}
\begin{align}
(P3):\, & \text{Find } \boldsymbol{\Theta}\\
&\text{Subject to, } \eqref{eq-null}, \eqref{eq-recip-const},\eqref{eq-loss-const}
\end{align} 

\end{subequations}

Such problem is non-convex and challenging to solve. Therefore, \gls{ao} is used to find $\boldsymbol{\Theta}$ that complies with the previously mentioned constraints by projecting a certain initial solution into $\mathcal{S}_{\mathrm{null}}$, $\mathcal{S}_{c_1}$ and $\mathcal{S}_{c_2}$ alternatively. Fortunately, the projections to $\mathcal{S}_{c_1}$ and $\mathcal{S}_{c_2}$ have been presented with closed-form solutions in \eqref{eq:proj-unim}, \eqref{eq:proj-resp}--\eqref{eq:proj-loss}, while the interference nulling projection follows \cite[Eq. (21a)]{Jiang2022-JSAC} such that
\begin{equation}
\Pi_{\mathcal{S}_\mathrm{null}}\left(\boldsymbol{\theta}_c\right) = \boldsymbol{\theta}_c - \bar{\mathbf{A}}_c^\ast\left(\bar{\mathbf{A}}_c^T\bar{\mathbf{A}}_c^{\ast}\right)^{-1}\bar{\mathbf{A}}_c^T\boldsymbol{\theta}_c\label{eq:proj-null}
\end{equation}

\begin{algorithm}[t]
\DontPrintSemicolon \KwInput{$\bar{\mathbf{A}}_c,\,G, \,\mathrm{Iter},\,\epsilon,\,\varepsilon$}
  \KwOutput{$\boldsymbol{\Theta}^{\star}$}
  Set $t=0$; Initialize $\boldsymbol{\Theta}^{(t)}$; Extract $\boldsymbol{\theta}_c^{(t)}$ \eqref{eq-theta-c} \;
  \If{G=1}{  $[\tilde{\boldsymbol{\theta}}_c^{(t)}]_{i}=\Pi_{\mathcal{S}_\mathrm{unim}}\left([{\boldsymbol{\theta}}_c^{(t)}]_{i}\right)$, $\forall i\in\{1,2,\ldots,N\}$\;
  $\tilde{\boldsymbol{\Theta}}^{(t)} = \mathrm{diag}\left(\tilde{\boldsymbol{\theta}}_c^{(t)}\right)$
  }
  \Else{
        \For{$g=:1:G$}{ $\boldsymbol{\Theta}_g^{(t)}\!=\!\mathrm{unvec}\!\left([\boldsymbol{\theta}_c^{(t)}]_{(g-1)N_g^2+1:gN_g^2},\,[N_g, N_g]\right)$\;
        $\tilde{\boldsymbol{\Theta}}_g^{(t)}=\Pi_{\mathcal{S}_\mathrm{sym}}\left(\Pi_{\mathcal{S}_\mathrm{uni}}\left(\boldsymbol{\Theta}_g^{(t)}\right)\right) $ \;
        }
        $\tilde{\boldsymbol{\theta}}_c^{(t)}=\left[\mathrm{vec}\left(\tilde{\boldsymbol{\Theta}}_1^{(t)}\right)^T,\ldots,\mathrm{vec}\left(\tilde{\boldsymbol{\Theta}}_G^{(t)}\right)^T\right]^T$\;
        $\tilde{\boldsymbol{\Theta}}^{(t)}=\mathrm{blkdiag}\left(\tilde{\boldsymbol{\Theta}}_1^{(t)},\ldots,\tilde{\boldsymbol{\Theta}}_G^{(t)}\right)$\;
  }
\For{$t=:1:\mathrm{Iter}$}{
$\boldsymbol{\theta}_c^{(t)}=\Pi_{\mathcal{S}_\mathrm{null}}\left(\tilde{\boldsymbol{\theta}}_c^{(t-1)}\right) $ \;
Repeat Lines $2$--$10$ \;
$\delta=\frac{\left\Vert\bar{\mathbf{A}}_c^T\tilde{\boldsymbol{\theta}}_c^{(t)}\right\Vert^2_2 -\left\Vert \bar{\mathbf{A}}_c^T\tilde{\boldsymbol{\theta}}_c^{(t-1)}\right\Vert^2_2}{\left\Vert \bar{\mathbf{A}}_c^T\tilde{\boldsymbol{\theta}}_c^{(t-1)}\right\Vert^2_2}$\;
\If{$\delta\leq\varepsilon$ or ${\scriptstyle{\left\Vert \bar{\mathbf{A}}_c^T\tilde{\boldsymbol{\theta}}_c^{(t)}\right\Vert^2_2}}<\epsilon$ or $t>\mathrm{Iter}$}{
\Break and \textbf{go to} Line $17$
}

}
$\boldsymbol{\Theta}^{\star} = \tilde{\boldsymbol{\Theta}}^{(t)}$

\caption{\gls{ao} for passive interference nulling.}
\label{alg:ao-nulling}
\end{algorithm}
\setlength{\textfloatsep}{5pt}

\begin{figure*}[t]
    \centering
    {\subfigure[Nulling Condition]{\includegraphics[scale=0.7]{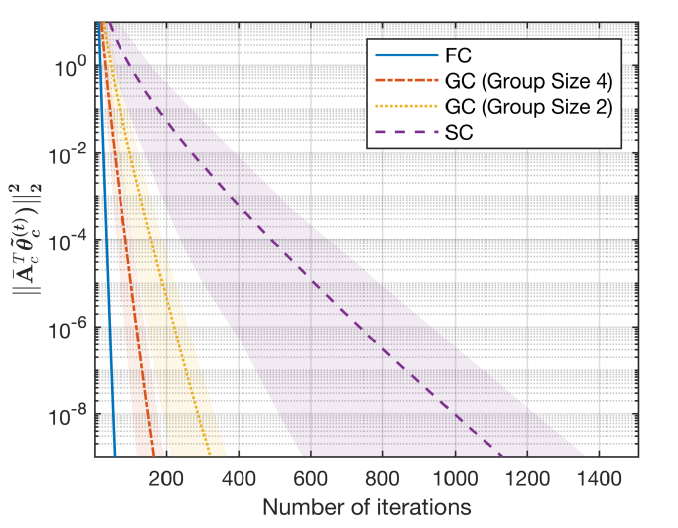}}}
    {\subfigure[Nulling Condition Relative Change]{\includegraphics[trim={0in 0 0 0 },clip,scale=0.7]{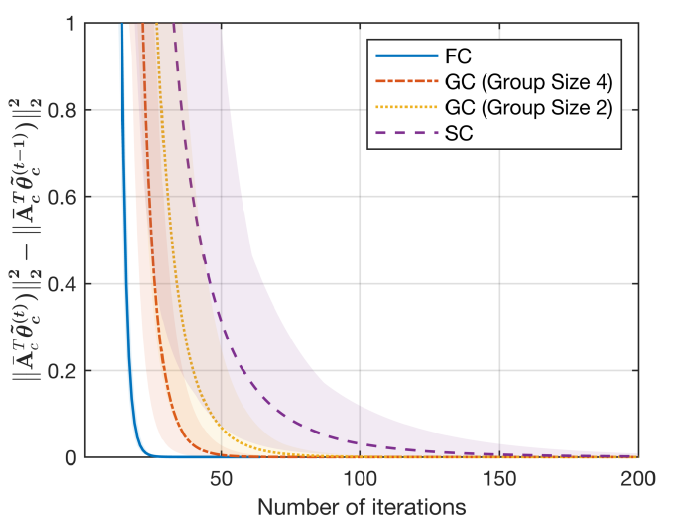}}}
    \caption{Convergence of Algorithm \ref{alg:ao-nulling} vs. number of iterations for passive interference nulling using fully-connected ``FC", group-connected ``GC", and single-connected ``SC" \glspl{bdris} with random initialization, where $K=8$ and $N=144$. a) Nulling condition $l^2$ norm. b) Nulling condition $l^2$ norm relative change.}
    \label{fig-res-converg}
\end{figure*}

\begin{table*}[]
\centering
\caption{Summary of the computational and circuit topology complexities for the proposed \gls{bdris} designs.}
\begin{tabular}{|ll||c||c|}
\hline
\multicolumn{2}{|c||}{\multirow{2}{*}{\gls{bdris} Design}}                                  & \multicolumn{1}{c||}{\multirow{2}{*}{Computational Complexity}} & \multicolumn{1}{c|}{Circuit Topology} \\
\multicolumn{2}{|c||}{}                                                                & \multicolumn{1}{c||}{}                                          & \multicolumn{1}{c|}{Complexity}       \\ \hline\hline

\multicolumn{1}{|l|}{\multirow{2}{*}{Single-Connected}} & Passive \gls{mrt}                  &    $\mathcal{O}\left(KN+N\right)$                   & \multirow{2}{*}{$N$}           \\ \cline{2-3}
\multicolumn{1}{|c|}{}                                  & Passive Interference Nulling &                    $\mathcal{O}\left(tN\!+K^6\!+K^4N\!+K^2N^2\!+K^2N\right)$      &                             \\ \hline
\multicolumn{1}{|l|}{\multirow{2}{*}{Fully-Connected}}  & Passive \gls{mrt}                  &   $\mathcal{O}\left(N^3+KN^2+N^2\right)$                       & \multirow{2}{*}{$\displaystyle\frac{N(1+N)}{2}$}           \\ \cline{2-3}
\multicolumn{1}{|l|}{}                                  & Passive Interference Nulling &                    $\mathcal{O}\left(tN^3\!+tN^2\!+K^6\!+K^4N^2\!+K^2N^4\!+K^2N^2\right)$      &                             \\ \hline
\multicolumn{1}{|l|}{\multirow{2}{*}{Group-Connected}}  & Passive \gls{mrt}                  &   $\mathcal{O}\left(GN_g^3+GKN_g^2+GN_g^2\right)$                       & \multirow{2}{*}{$\displaystyle\frac{N(1+N_g)}{2}$}           \\ \cline{2-3}
\multicolumn{1}{|l|}{}                                  & Passive Interference Nulling &                    $\mathcal{O}\left(tGN_g^3\!+tGN_g^2\!+K^6\!+K^4GN_g^2\!+K^2G^2N_g^4\!+K^2GN_g^2\right)$      &                             \\ \hline
\end{tabular}
\label{tab-complexity}
\end{table*} 

A pseudo-code that illustrates the steps to obtain $\boldsymbol{\Theta}^\star$ through \gls{ao} is given in Algorithm \ref{alg:ao-nulling}. The algorithm needs the interference channels matrix $\bar{\mathbf{A}}_c$ and the number of groups $G$ as inputs. It starts by initializing the scattering matrix, which can be either a random initialization or a specific initialization. In addition, the initialized scattering matrix is projected to $\mathcal{S}_{c_1}$ and $\mathcal{S}_{c_2}$ in Lines $2$--$10$. Lines $11$--$16$ illustrate the iterative \gls{ao} which projects the $\boldsymbol{\Theta}$ from the previous iteration to interference nulling set and $\mathcal{S}_{c_1}$ and $\mathcal{S}_{c_2}$. This continues for a certain number of iterations until the breaking condition on line $15$ is met. On line $14$, $\delta$ is computed which is the $l^2$ norm for the nulling condition difference between the current iteration and the previous iteration. Finally, $\boldsymbol{\Theta}^{\star}$ that achieves interference nulling and belongs to $\mathcal{S}_{c_1}$ and $\mathcal{S}_{c_2}$ is returned. It is worth noting that $\epsilon$ and $\varepsilon$ determine the algorithm convergence speed, where the former is the relative rate of change tolerance and the latter is the nulling condition norm tolerance. 

Fig. \ref{fig-res-converg} compares the convergence of Algorithm \ref{alg:ao-nulling} for fully-connected, group-connected and single-connected \glspl{bdris}. Random initialization is assumed for both, with $N=144$, $K=8$, $\epsilon=10^{-10}$, $\varepsilon=10^{-6}$, $\mathrm{Iter}=10^4$, and $100$ realizations are considered for Monte-Carlo simulation. Also, \gls{iid} Rayleigh fading channels are considered. The curves represent the average convergence of the algorithm over $10^4$ iterations, while the shades represent the possible variations, i.e., the maximum and minimum convergence. It can be observed that fully-connected \gls{bdris} achieves a rapid interference nulling, reaching $10^{-8}$ with significantly fewer iterations compared to single-connected \gls{bdris}. In addition, the relative change for the fully-connected \gls{bdris} stabilizes after fewer iterations compared to the single-connected \gls{bdris}. Moreover, the the fully-connected \gls{bdris} maintains a stable pattern, effectively achieving interference nulling, unlike the single-connected \gls{bdris} which is more dependent on the initialization of the scattering matrix, resulting in less stable interference nulling. Furthermore, the behavior of the group-connected \gls{bdris} falls between the two extremes. Specifically, as the group size approaches 
$N$, the performance becomes closer to that of a fully-connected \gls{bdris}. Conversely, as the group size approaches $1$, the performance becomes similar to that of the single-connected \gls{bdris}.

\textit{Computational Complexity}: The computational complexity of Algorithm \ref{alg:ao-nulling} is dominated by computing the projections inside the loop. For instance, in \eqref{eq:proj-null}, its main part, i.e., $\bar{\mathbf{A}}_c^\ast\left(\bar{\mathbf{A}}_c^T\bar{\mathbf{A}}_c^{\ast}\right)^{-1}\bar{\mathbf{A}}_c^T$, can be computed outside the loop as it only depends on the interference channels matrix. Hence, its computational complexity can be written as $\mathcal{O}\left(K^6+K^4GN_g^2+K^2G^2N_g^4+K^2GN_g^2\right)$. For the single-connected architecture, the unit-modulus projection has a computational complexity of $\mathcal{O}\left(tN\right)$, while the symmetric and unitary projections associated with group-connected architecture have a computational complexity of $\mathcal{O}\left(tGN_g^3\right)$ and $\mathcal{O}\left(tGN_g^2\right)$. Consequently, the overall algorithm's computational complexity is $\mathcal{O}\left(tGN_g^3\!+tGN_g^2\!+K^6\!+K^4GN_g^2\!+K^2G^2N_g^4\!+K^2GN_g^2\right)$. Typically, when $N\gg K$ the computational complexity can be written as $\mathcal{O}\left(K^2G^2N_g^4\right)$.
Moreover, the Numerical Results and Discussions Section will show that initializing based on passive \gls{mrt} solution, \eqref{eq-mrt-final-SC}, \eqref{eq:mrt-final-GC}, improves the sum rate performance compared to the random initialization. 

\textit{Channel State Information Requirements}: The interference channels matrix $\bar{\mathbf{A}}_c$ is needed to compute $\boldsymbol{\Theta}^\star$. This can be found by pre-designing the variation of $\mathbf{\Theta}$ for the single/group-connected \gls{bdris} architecture to be further used in the training process \cite{Hongyu2024-ChannelEst}.

For the readers convenience, Table \ref{tab-complexity} summarizes the computational complexity and circuit toplogy complexity for the different \gls{bdris} passive \gls{mrt} and passive interference nulling designs.

\section{Stage 2: BS Power Allocation/Precoding}\label{sec-stage2}
As $\boldsymbol{\Theta}^{\star}$ is determined at stage 1, the precoding matrix $\mathbf{P}$ relies only on the equivalent channel matrix, i.e., $\mathbf{E}=\mathbf{H}\boldsymbol{\Theta}^{\star}\mathbf{W}$ which is of low-dimension compared to the individual channel matrices. Hence, \eqref{eq-main-opt} can be written as 
\begin{subequations}
\begin{align}
(P4)\!:\, & \!\max_{\mathbf{P}}\!\!\sum_k\log_2\!\!\left(\!1\!+\!\frac{\left\vert \mathbf{h}_{k}^{T}\boldsymbol{\Theta}^{\star}\mathbf{W}\mathbf{p}_k\right\vert ^{2}}{\sum_{i\neq k}\left\vert \mathbf{h}_{k}^{T}%
\boldsymbol{\Theta}^{\star}\mathbf{W}\mathbf{p}_i\right\vert ^{2}+N_0}\!\right)\\
&\text{Subject to, } \eqref{eq-fro-const}
\end{align}
\end{subequations}
Solving this optimization problem for the general case is not straightforward. Therefore, a distinction is made between the passive \gls{mrt} and interference nulling solutions. 

\subsection{Power Allocation/Precoding for Passive MRT}
Since the passive \gls{mrt} presented in Sec. \ref{sec-stage1}-\ref{subsec-MRT} does not set the interference terms to zero, a low-complexity precoder can be considered at the \gls{bs} such as \gls{zf}. Consequently, 
\begin{equation}
    \mathbf{P}=\mathbf{E}^{\dag}\sqrt{\frac{P_{\max}}{\mathrm{Tr}\left(\mathbf{E}^{\dag H}\mathbf{E}\right)}} \label{eq-bs-zf}
\end{equation}
where the normalization factor ensures $\left\Vert\mathbf{P}\right\Vert^2_F=P_{\max}$. The computational complexity of the \gls{zf} solution is bounded by the dimension of the equivalent channels matrix, and hence, it can be written as $\mathcal{O}\left(K^3\right)$.

Alternatively, employing simple power allocation approaches at the \gls{bs} will ease the requirements of high resolution digital-to-analog converters \cite{SIM2023-ICC}. In addition, the number of real variables will reduce to $K$, unlike the full precoder which has $K^2$ complex variables. Two power allocation approaches are devised, which are the uniform power and optimized power. The former considers equal power allocation among the signal vector elements, i.e., 
\begin{equation}
\mathbf{P}=\mathrm{diag}\left(\sqrt{p_1},\sqrt{p_2},\ldots,\sqrt{p_K}\right) \label{eq-upa}
\end{equation}
such that $p_k = \frac{P_{\max}}{K}$, $\forall k$. Hence, it is agnostic to the equivalent channels matrix. On the other hand, while the equivalent channels matrix is known at the \gls{bs}, the optimized power can be designed to maximize the sum rate of the system. Therefore, the sum rate maximization problem can be formulated as \cite[Eq. (3)]{Huang2019-TWC},
\begin{subequations}
\begin{equation}
(P5)\!:\,\!\max_{p_k}\!\!\sum_k\log_2\!\!\left(\!1\!+\!\frac{p_k\left\vert \mathbf{h}_{k}^{T}\boldsymbol{\Theta}^{\star}\mathbf{w}_{k}\right\vert ^{2}}{\sum_{i\neq k}p_i\left\vert \mathbf{h}_{k}^{T}%
\boldsymbol{\Theta}^{\star}\mathbf{w}_{i}\right\vert ^{2}\!\!+N_0}\!\right)\label{eq-opt-pc1}
\end{equation}
\begin{align}
\text{Subject to, } &&
\sum_k p_k &= P_{\max}\label{eq-opt-pc2}\\
&& p_k &\geq 0 \label{eq-opt-pc3}
\end{align}
\end{subequations}
where \eqref{eq-opt-pc1} is non-linear and non-convex, \eqref{eq-opt-pc2} is affine and linear and \eqref{eq-opt-pc3} defines a hyperplane and is convex. Such optimization problem can be solved efficiently using interior-point optimization algorithms.

\subsection{Power Allocation for Passive Interference Nulling}

Since the passive interference nulling presented in Sec. \ref{sec-stage1}-\ref{subsec-intNull} achieves \gls{zf} when sufficient number of \glspl{re} exist, i.e., $N\geq\left\lceil\frac{4K(K-1)}{1+N_g}\right\rceil$, $\mathbf{E}$ is a diagonal matrix. Hence, the transmission results in $K$ virtual parallel streams. Therefore, full precoding is not required to maximize the sum rate. Alternatively, water-filling power allocation can be used, which is optimal for \eqref{eq-opt-pc1}. The closed-form water-filling solution is given as
\begin{equation}
    p_k^{\star} = \left(\frac{1}{\alpha} - \frac{N_0}{\left\vert \mathbf{h}_{k}^{T}%
\boldsymbol{\Theta}^{\star}\mathbf{w}_k\right\vert}\right)^{+} \label{eq-wf}
\end{equation}
where $\alpha$ is the dual variable that can be found using bi-section search. It is worth noting that the water-filling solution for passive interference nulling converges to the uniform power solution at high \glspl{snr}, unlike passive \gls{mrt} which still requires power allocation/precoding at the \gls{bs}. Hence, in the presence of imperfect \gls{csi} at high \glspl{snr}, passive interference nulling is more robust as \gls{csi} is only required at the \gls{bdris}, compared to passive \gls{mrt} which requires \gls{csi} at the \gls{bs} and \gls{bdris}.

\section{Numerical Results and Discussions}\label{sec-results}
This section presents the numerical results for the proposed two-stage designs as well as the benchmarks. 

\subsection{Simulation Setup}
The adopted channel model is \gls{iid} Rayleigh fading with the distance-dependent large-scale fading given as $\Upsilon\left(d\right)=C_0\left(\frac{d}{d_0}\right)^{-\rho}$, where $d$ denotes the distance, $C_0=-30$ dB denotes the reference pathloss, $d_0=1$ m is reference distance and $\rho=2.2$ is the pathloss exponent. We consider the system to be operating at $2.4$ GHz, with noise power spectral density $N_0=-80$ dBm that is identical for all users. Unless otherwise stated, we consider the distance between \gls{bs} and \gls{bdris} to be $50$ m and the distance between the \gls{bdris} and users to be $2.5$ m. We assume the channel state information to be known perfectly at the \gls{bs}, \gls{bdris} and the users. For Algorithm \ref{alg:ao-nulling}, $\epsilon=10^{-6}$, $\varepsilon=10^{-6}$ and $\mathrm{Iter}=10^2$ are set. Hundred realizations are considered for Monte-Carlo simulation. The simulations were conducted on a work station equipped with $3.49$ GHz $64$ bits Apple M2 Pro chip with 12‑core CPU and $16$ GB RAM.

\subsection{Notations for Proposed Designs and Benchmarks}\label{subsec:benchmarks}
Various \textit{benchmarks} are used to compare the performance of our \textit{proposed} designs for the \gls{bdris}. The notations for the stage 1 \textit{proposed} designs are given as follows:
\begin{itemize}
    \item \textbf{MRT} denotes the passive \gls{mrt} solution introduced in Sec. \ref{sec-stage1}-\ref{subsec-MRT}.
    \item \textbf{Null} denotes the passive interference nulling introduced in Sec. \ref{sec-stage1}-\ref{subsec-intNull}, where \gls{mrt} initializations \eqref{eq-mrt-final-SC}, \eqref{eq:mrt-final-GC} are considered for Algorithm \ref{alg:ao-nulling}.  
\end{itemize}
The notations for the stage 2 processes at the \gls{bs} are: 
\begin{itemize}
    \item \textbf{ZF} denotes the zero-forcing precoder shown in \eqref{eq-bs-zf}.
    \item \textbf{UP} denotes the uniform power allocation shown in \eqref{eq-upa}.
    \item \textbf{RM} denotes the rate maximization optimization $(P5)$ shown in \eqref{eq-opt-pc1}--\eqref{eq-opt-pc3} and the closed-form water-filling shown in \eqref{eq-wf}.
\end{itemize}
The considered \textit{benchmarks} are:
\begin{itemize}
    \item \textbf{Joint} denotes the joint-design for the \gls{bdris} following \cite{Hongyu2023-JSAC}, which has a computational complexity of $\mathcal{O}(t_1t_2t_3GN_g^3)$ for the group-connected, where $t_1$, $t_2$ and $t_3$ are the number of iterations for the algorithm loops.
    \item \textbf{Max-F} denotes the Frobenius norm-based maximum channel gain design for the \gls{bdris} scattering matrix. It aims to $\max_{\boldsymbol{\Theta}}\left\Vert\mathbf{E}\right\Vert^2_F$, which has a relaxed closed-form solution that follows \cite{Demir2022-TWC}. The optimal solution that belongs to $\mathcal{S}_{\mathrm{SC}_1}$ and $\mathcal{S}_{\mathrm{SC}_2}$ for the single-connected is given as 
    \begin{equation}
    \left[\boldsymbol{\Theta}^\star\right]_{i,j} = \left\{\begin{array}{lr}
        \Pi_{\mathcal{S}_\mathrm{unim}}\left(\left[\mathbf{v}_{\max}\right]_{i}\right) ,& i=j\\
        0 ,& i\neq j
        \end{array}
        \right.
\end{equation}
 where $\mathbf{v}_{\max}$ is the Eigenvector of $\mathbf{A}_c^H\mathbf{A}_c$ associated with its dominant Eigenvalue. 
    The optimal solution that belongs to $\mathcal{S}_{\mathrm{GC}_1}$ and $\mathcal{S}_{\mathrm{GC}_2}$ for the group-connected is given as \eqref{eq:mrt-final-GC1} and 
    \begin{equation}
    \boldsymbol{\Theta}_g^{\star}=\Pi_{\mathcal{S}_\mathrm{symuni}}\left(\!\mathrm{unvec}\left(\!\sqrt{N_g}\mathbf{v}_{g,\max},
[N_g,N_g]\!\right)\!\right)
    \end{equation}
      where $\mathbf{v}_{g,\max}=\left[\mathbf{v}_{\max}\right]_{(g-1)N_g^2+1:gN_g^2}$. Its computational complexity comes from computing \gls{svd} for $GN_g^2\times GN_g^2$ matrix and is $\mathcal{O}(G^3N_g^6)$. At the \gls{bs}, \gls{zf} is performed following \eqref{eq-bs-zf}, which has a computational complexity of $\mathcal{O}(K^3)$ that comes from computing the Moore-pseudoinverse of $\mathbf{E}$.
    \item \textbf{Spec-ZF} denotes the the specular reflection by \gls{bdris}, i.e., $\boldsymbol{\Theta}^{\star}=\mathbf{I}$, where the \gls{bs} performs \gls{zf} based on $\mathbf{E}$ such that $\mathbf{P}=\mathbf{E}^{\dag}\sqrt{\frac{P_{\max}}{\mathrm{Tr}\left(\mathbf{E}^{\dag H}\mathbf{E}\right)}}$ and has a computational complexity of $\mathcal{O}(K^3)$. 
    \item \textbf{Spec-MRT} denotes the the specular reflection by \gls{bdris}, i.e., $\boldsymbol{\Theta}^{\star}=\mathbf{I}$, where the \gls{bs} performs \gls{mrt} based on $\mathbf{E}$ such that $\mathbf{P}=\mathbf{E}^{H}\sqrt{\frac{P_{\max}}{\mathrm{Tr}\left(\mathbf{E}^H\mathbf{E}\right)}}$ and has a computational complexity of $\mathcal{O}(K^2)$. 
\end{itemize}

\begin{figure}[t]
    \centering
    {\includegraphics[trim={0in 0 0 0 },clip,scale=0.7]{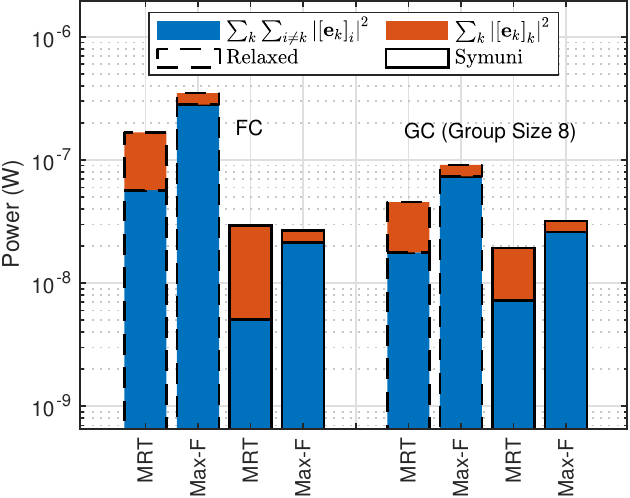}}
    \caption{Desired and interference powers for fully-connected ``FC" and group-connected ``GC" \gls{bdris} considering \textbf{MRT} and \textbf{Max-F} for $p_k=1,\forall k$, $K=5$ and $N=16$.}
    \label{fig-symuni_relaxed-bar}
    \end{figure}

\begin{figure*}[t]
    \centering
    {\subfigure[]{\includegraphics[trim={0in 0 0 0 },clip,scale=0.7]{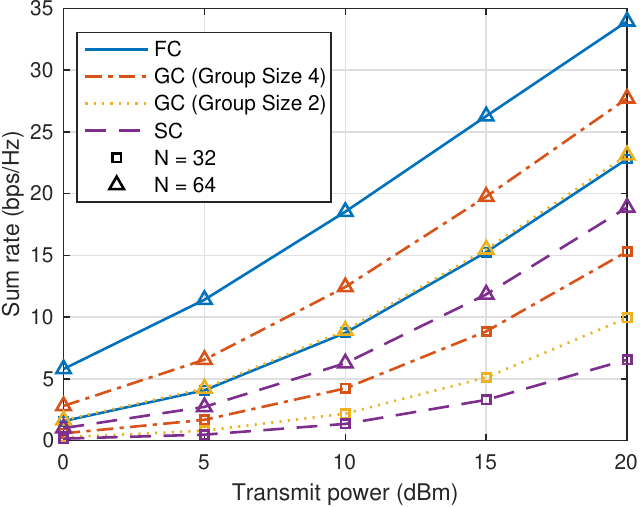}}}
    {\subfigure[]{\includegraphics[trim={0in 0 0 0 },clip,scale=0.7]{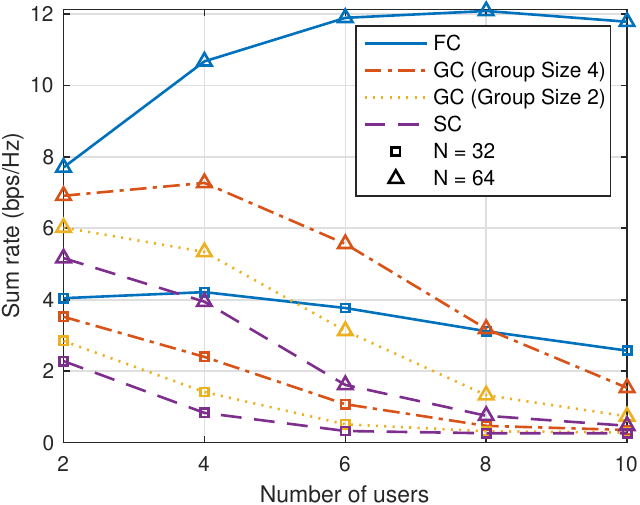}}}
    \caption{Sum rate for passive interference nulling using fully-connected ``FC", group-connected ``GC", and single-connected ``SC" \glspl{bdris} with \gls{mrt} initialization, where a) x-axis is transmit power and $K=5$. b) x-axis is number of users and $P_{\max}=5$ dBm.}
    \label{fig-NULL-sRxPt}
    \end{figure*}

\begin{figure*}[t]
    \centering
    {\subfigure[Passive MRT]{\includegraphics[trim={0in 0 0 0 },clip,scale=0.7]{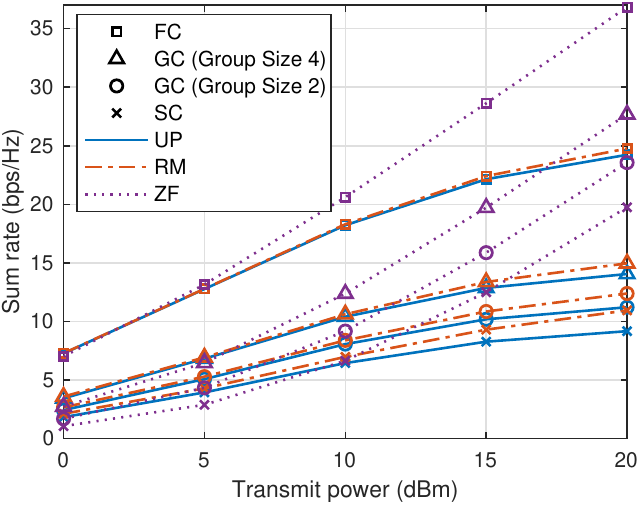}}}
    {\subfigure[Passive Interference Nulling]{\includegraphics[trim={0in 0 0 0 },clip,scale=0.7]{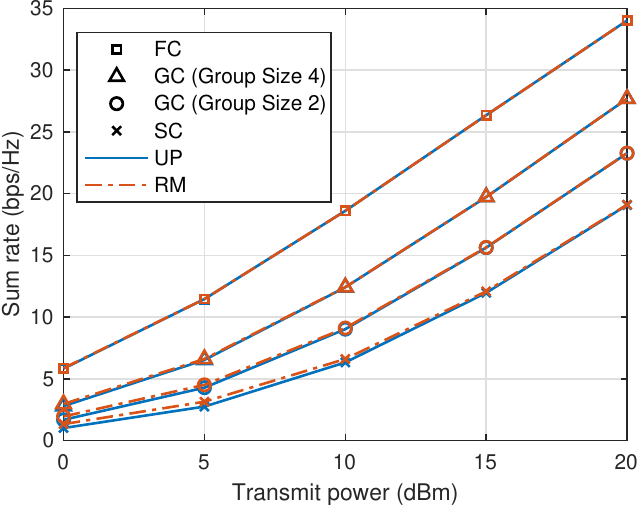}}}
    \caption{Sum rate vs. transmit power for the stage 2 processes at the \gls{bs} using fully-connected ``FC", group-connected ``GC", and single-connected ``SC" \glspl{bdris}, where $K=5$, $P_{\max}=5$ dBm, $N=64$. a) Passive \gls{mrt}. b) Passive interference nulling.}
    \label{fig-Proposed-sRxPt}
    \end{figure*}

\subsection{Simulation Results}
Fig. \ref{fig-symuni_relaxed-bar} illustrates the received \gls{sinr} at stage 1 of a fully-connected and group-connected \glspl{bdris} with $K=5$, comparing relaxed and symmetric unitary solutions for \textbf{MRT} and \textbf{Max-F}. Specifically, the figure breakdowns \eqref{eq-stage1-SINR} into the sum of desired signals' powers and sum of interference signals' powers for $N=16$ in a stacked bar chart. It is seen that the relaxed \textbf{Max-F} achieves the highest Frobenius norm having the highest bar. Nonetheless, the symmetric unitary solution for \textbf{Max-F} does not guarantee the maximization of the Frobenius norm as seen for the fully-connected case, where the symmetric unitary \textbf{MRT} is better. In addition, the sum of desired signals' powers for the relaxed and symmetric unitary \textbf{MRT} are higher than that of \textbf{Max-F} as expected, whereas the sum of interference signals' powers is higher for \textbf{Max-F} compared to \textbf{MRT} as the latter only maximizes the diagonal entries of the equivalent channel but the former maximizes all entries. Such observation indicates that \textbf{MRT} would have a better \gls{sinr} at the users compared to \textbf{Max-F}. Furthermore, although the \gls{svd}-based symmetric unitary projection degrades the sum of desired signals' powers for both \textbf{Max-F} and \textbf{MRT}, it reduces the sum of interference signals' powers for \textbf{MRT} at a higher rate such that the \gls{sinr} can be effectively improved, unlike \textbf{Max-F} which reduces both the sum of signal powers and sum of interference at the same rate.

    \begin{figure*}[t]
    \centering
    {\subfigure[Number of users]{\includegraphics[scale=0.7]{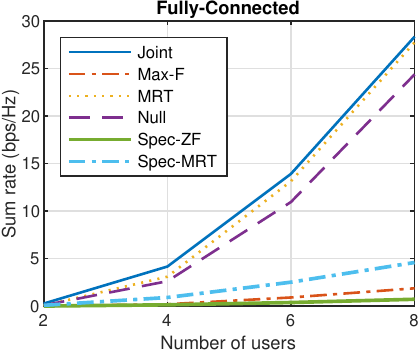}
    \includegraphics[scale=0.7]{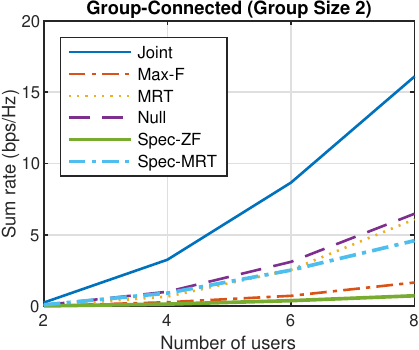}
    \includegraphics[scale=0.7]{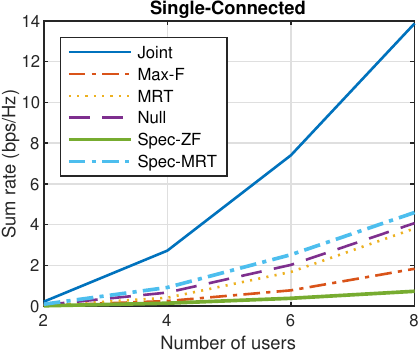}}}
    {\subfigure[Number of \glspl{re}]{\includegraphics[scale=0.7]{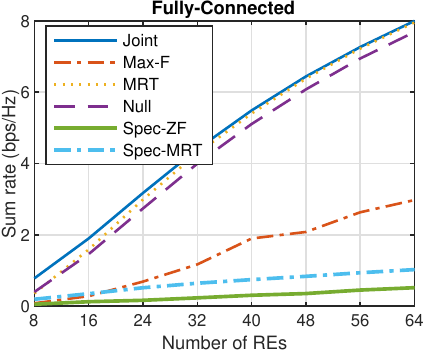}
    \includegraphics[scale=0.7]{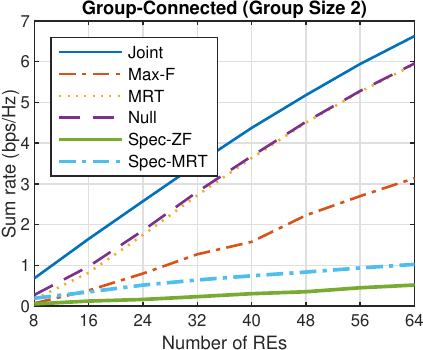}
    \includegraphics[scale=0.7]{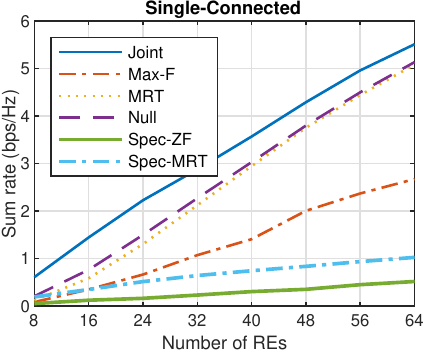}}}
    \caption{Sum rate for the \textit{benchmarks} and \textit{proposed} designs for fully-connected, group-connected and single-connected \glspl{bdris} with $P_{\max}=5$ dBm. a) x-axis is number of users, $N=2K(K-1)$. b) x-axis is number of \glspl{re}, $K=2$.}
    \label{fig-res-sumrate}
\end{figure*}

Fig. \ref{fig-NULL-sRxPt} compares the sum rate performance of interference nulling using single/group/fully-connected \glspl{bdris}. The system uses \gls{mrt} initialization and uniform power allocation at the \gls{bs}, with $K=5$ and $N\in\{32,64\}$\footnote{Note that uniform power allocation is not optimal to maximize the sum rate. Nonetheless, it is considered in this figure to assess the sum rate performance of passive interference nulling using \gls{bdris}.}. As shown in Fig. \ref{fig-NULL-sRxPt}a, the fully-connected \gls{bdris} outperforms the single-connected \gls{bdris} for both $N=32$ and $N=64$ with a huge gap. This performance gap is linked to the fully-connected \gls{bdris} requiring fewer \glspl{re} to effectively null the interference with high probability. Specifically, \eqref{eq-N-bound-null} shows that the fully-connected \gls{bdris} requires $N\geq2K-1$, while the single-connected \gls{bdris} needs $N\geq2K(K-1)$. Furthermore, Fig. \ref{fig-NULL-sRxPt}b shows that increasing the number of users, while fixing the number of \glspl{re}, does not necessarily improve the sum rate. For instance, when $N$ is sufficiently large, the fully-connected \gls{bdris} has an improved sum rate as more users are served. This occurs because, when $N\gg2K-1$, interference is nulled successfully using the required $2K-1$ \glspl{re}, leaving a substantial potion of the remaining \glspl{re} to enhance the signal power. In contrast, both the group-connected and single-connected \glspl{bdris} show a degraded sum rate as the number of users increases. This degradation arises because $N$ is not sufficiently large to achieve $N\gg2K(K-1)$ and $N\gg\left\lceil\frac{4K(K-1)}{1+N_g}\right\rceil$. Consequently, most of the \glspl{re} are used to null the interference, i.e., $2K(K-1)$ and $\left\lceil\frac{4K(K-1)}{1+N_g}\right\rceil$ \glspl{re}, leaving only a small portion of \glspl{re} to enhance the signal power.
    
Fig. \ref{fig-Proposed-sRxPt} shows the sum rate of single/group/fully-connected \glspl{bdris} as a function of transmit power at the \gls{bs}. It considers the \textit{proposed} \gls{bdris} designs, specifically considering the \textbf{MRT} and \textbf{Null} schemes. Additionally, the impact of different stage 2 processes at the \gls{bs} is considered. The highlighted scenario in the figure is $K=5$ users and $N=64$ \glspl{re}. Focusing on \textbf{UP}, Fig. \ref{fig-Proposed-sRxPt} shows that \textbf{MRT} outperforms \textbf{Null} at low \glspl{snr}, where performance performance is primarily dominated by noise. This is because \textbf{MRT} provides a better channel alignment compared to \textbf{Null}. However, at moderate and high \glspl{snr}, \textbf{MRT} becomes interference-limited leading its performance to saturate, unlike \textbf{Null}, which inherently eliminates the interference. On the other hand, \textbf{RM} power allocation schemes improve the sum rate of \textbf{MRT} in general. While \textbf{RM} also enhances the sum rate of \textbf{Null} at low \glspl{snr}, the improvement at high \glspl{snr} is negligible. Finally, considering the fully-connected \gls{bdris} as an example, \textbf{MRT} with \textbf{ZF} precoding at the \gls{bs} outperforms \textbf{Null} with \textbf{RM} at the \gls{bs} by $1.3$ dB at $20$ bps/Hz. Thus, \textbf{MRT} with \textbf{ZF} at the \gls{bs} achieves a balance between low computational complexity and good sum rate performance at low and moderate \glspl{snr}, while offering the best sum rate performance at high \glspl{snr}.

Fig. \ref{fig-res-sumrate} presents the sum rate performance for both the \textit{proposed} designs and the \textit{benchmarks} across single/group/fully-connected \glspl{bdris}, where $P_{\max}=5$ dBm. In Fig. \ref{fig-res-sumrate}a, the sum rate is illustrated as a function of number of users, where $N=2K(K-1)$ \glspl{re}. It can be seen that \textbf{Joint} serves as an upper bound, with the gap between the upper bound and the \textit{proposed} designs varying based on the \gls{bdris} architecture. For instance, the gap is extremely tight for the fully-connected \gls{bdris} since $N\gg2K-1$, meaning that $N$ is sufficiently large compared to the bound in \eqref{eq-N-bound-null}. In contrast, the group-connected and single-connected \glspl{bdris} exhibit a much larger gap from the upper bound, as $N$ is not sufficiently large when compared to the bound in \eqref{eq-N-bound-null}. For instance, at $K=8$, the fully-connected \gls{bdris} with \textbf{Joint} achieves a sum rate of $28.3$ bps/Hz, whereas \textbf{MRT} with \gls{zf} at the \gls{bs} achieves a sum rate of $27.7$ bps/Hz\footnote{It is worth noting that \textbf{Joint} only considers \eqref{eq-loss-const} constraint, while the rest \gls{bdris} designs consider both \eqref{eq-recip-const} and \eqref{eq-loss-const} constraints.}. Additionally, both \textit{proposed} \gls{bdris} designs outperform \textbf{Max-F} designs because the \textit{proposed} \gls{bdris} designs have better interference management in multiuser systems. In addition, \textbf{Spec-ZF} and \textbf{Spec-MRT} are both performing poorly, particularly for the fully-connected \gls{bdris}. This is because these designs fail to leverage the \gls{bdris} to mitigate the poor channel gains caused by the double-fading effect, making reliance on the \gls{bs} solely for precoding inefficient. For instance, \gls{zf} wastes most of the power for channel inversion leaving very small power gains. Moreover, the performance of the \textit{proposed} designs improve as the number of users/\glspl{re} increase which is justified by the growing Frobenius norm of the scattering matrix, indicating their effectiveness in handling dense multiuser scenarios. While the gap between \textbf{MRT} and \textbf{Null} with \gls{mrt} initialization increases as the number of users increase, the latter can be a good candidate to achieve a reasonable sum rate performance meanwhile maintaining a negligible complexity at the \gls{bs}.

Furthermore, Fig. \ref{fig-res-sumrate}b shows the sum rate as a function of number of \glspl{re} for $K=2$. It is seen that \textbf{Joint} serves as an upper bound here too, where the gap between the upper bound and the \textit{proposed} designs is very tight given that the range of $N$ is sufficiently large for all \gls{bdris} architectures given $K=2$. This indicates that passive \gls{mrt} with \gls{zf} precoding at the \gls{bs} achieves local optimality while maintaining low complexity, and is considered a good candidate for \gls{bdris} with many \glspl{re}. The specular reflection \textbf{BD-RIS} curves on the other hand are both lower bounds and the justification holds from Fig. \ref{fig-res-sumrate}a. Also, the significance of \textbf{Max-F} is shown for high number of \glspl{re}. Their poor performance with relatively small $N$ is justified by the fact that the equivalent channel is ill-conditioned/poorly conditioned for such $N$ values and it improves as $N$ increases.

\begin{figure}[t]
    \centering
    {{{\includegraphics[scale=0.7]{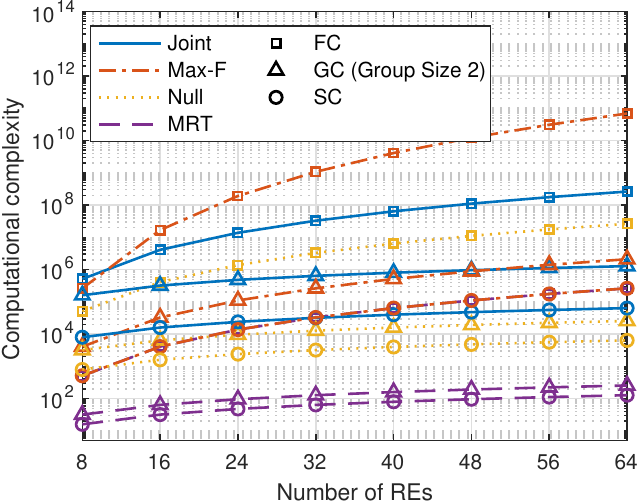}}}}
    \caption{Computational complexity as a function of number of \glspl{re}. The \textit{proposed} designs and \textit{benchmarks} are considered for fully-connected ``FC", group-connected ``GC" and single-connected ``SC" \glspl{bdris}, where $K=2$.}
    \label{fig-time-complex}
\end{figure}

Fig. \ref{fig-time-complex} shows the computational complexity considering the \textit{proposed} designs and \textit{benchmarks} for single/group/fully-connected, where $K=2$. The worst-case scenario is considered for \textbf{Joint} and \textbf{Null} with \gls{mrt} initialization, such that $t_1=50$, $t_2=20$, $t_3=5$, and $t=100$. As presented in Table \ref{tab-complexity} and Sec. \ref{sec-results}-\ref{subsec:benchmarks}, \textbf{MRT} has the lowest time complexity while the \textbf{Max-F} has the highest. Although \textbf{Joint} has a cubic time complexity, the complexity from the iterative algorithm is not negligible. Furthermore, \textbf{Null} has a lower complexity compared to both \textit{benchmarks}.

\section{Conclusions and Future Work}\label{sec-concl}
To conclude, this work proposed a passive multiuser beamforming for single/group/fully-connected \gls{bdris}-enabled \gls{mumiso} systems. The passive multiuser beamforing designs are inspired by \gls{mrt} and interference nulling. The former passive beamforming is presented in closed-form while the latter is presented in an \gls{ao} framework. At the \gls{bs} side, optimized/uniform power allocation and active precoding are considered. It is seen that the fully-connected \gls{bdris} reduces the number of \glspl{re} required to achieve interference nulling and achieves faster convergence compared to \gls{dris}. In addition, the passive \gls{mrt} at \gls{bdris} with \gls{zf} precoding at the \gls{bs} presents an effective low-complexity solution that simplifies channel estimation and achieves local optimality for \gls{mumiso} scenarios with many \glspl{re}.

For the future work, we will consider deriving a closed-form unitary symmetric scattering matrix to achieve \textit{interference nulling} for the \gls{bdris}. In addition, we will analyze the average \gls{sinr} and outage probability for the \textit{passive \gls{mrt}}. An extension to multi-antenna users will also be considered. Furthermore, we will generalize the \gls{bdris} scattering matrix design and \gls{bs} precoder design to account for the imperfect \gls{csi}.
\bibliographystyle{IEEEtran}

\end{document}

%% file: abbrev.tex
\usepackage[acronym,nonumberlist,style=super,toc]{glossaries}

\makeatletter
\newcommand*{\glsplainhyperlink}[2]{%
  \colorlet{currenttext}{.}
  \colorlet{currentlink}{\@linkcolor}
  \hypersetup{linkcolor=currenttext}
  \hyperlink{#1}{#2}%
  \hypersetup{linkcolor=currentlink}
}
\let\@glslink\glsplainhyperlink
\makeatother

\newacronym{5g}{5G}{fifith generation}
\newacronym{6g}{6G}{sixth generation}

\newacronym{awgn}{AWGN}{additive white Gaussian noise}
\newacronym{ao}{AO}{alternating optimization}

\newacronym{bs}{BS}{base station}
\newacronym{bdris}{BD-RIS}{beyond diagonal reconfigurable intelligent surfaces}

\newacronym{csi}{CSI}{channel state information}
\newacronym{cdf}{CDF}{cumulative  distribution  function}

\newacronym{dl}{DL}{downlink}
\newacronym{dris}{D-RIS}{diagonal reconfigurable intelligent surfaces}

\newacronym{iid}{i.i.d}{independent and identically distributed}
\newacronym{iot}{IoT}{Internet of Things}
\newacronym{irs}{IRS}{intelligent reflecting surfaces}
\newacronym{isac}{ISAC}{integrated sensing and communication}

\newacronym{jmld}{JMLD}{joint-multiuser maximum likelihood detector}

\newacronym{lut}{LUT}{look-up table}
\newacronym{lhs}{LHS}{left-hand side}

\newacronym{mpsk}{$M$-PSK}{M-ary phase shift keying}
\newacronym{mimo}{MIMO}{multiple-input-multiple-output}
\newacronym{miso}{MISO}{multiple-input-single-output}

\newacronym{mumiso}{MU-MISO}{multiuser multiple-input-single-output}

\newacronym{mld}{MLD}{maximum likelihood detector}
\newacronym{mrt}{MRT}{maximum ratio transmission}

\newacronym{ndris}{ND-RIS}{non-diagonal reconfigurable intelligent surfaces}

\newacronym{pc}{PC}{power control}

\newacronym{qos}{QoS}{quality of service}

\newacronym{ris}{RIS}{reconfigurable intelligent surfaces}
\newacronym{re}{RE}{reflecting element}
\newacronym{rf}{RF}{radio-frequency}
\newacronym{rhs}{RHS}{right-hand side}

\newacronym{svd}{SVD}{singular value decomposition}

\newacronym{se}{SE}{spectral efficiency}
\newacronym{snr}{SNR}{signal to noise ratio}
\newacronym{siso}{SISO}{single-input-single-output}
\newacronym{sinr}{SINR}{signal to interference and noise ratio}
\newacronym{swipt}{SWIPT}{simultaneous wireless information and power transfer}

\newacronym{ul}{UL}{uplink}
\newacronym{uav}{UAV}{unmanned aerial vehicle}

\newacronym{zf}{ZF}{zero forcing}